\documentclass[aps,prc,twocolumn,showpacs,superscriptaddress]{revtex4}  
\usepackage{amssymb,amsmath,mathtools,color}
\usepackage{graphicx}
\usepackage{color}

\usepackage{epsfig}
\usepackage{bm}

\graphicspath{%
    {converted_graphics/}
    {/}
}

\begin{document}

\title{Three-body Faddeev equations in two-particle Alt-Grassberger-Sandhas form with  Distorted-Wave-Born-approximation amplitudes as effective potentials}
\author{A. M. Mukhamedzhanov}
\affiliation{Cyclotron Institute, Texas A$\&$M University, College Station, TX 77843, USA}

\begin{abstract}
$A(d,p)B$ reactions on heavier nuclei are peripheral at sub-Coulomb energies and can be peripheral at energies above the Coulomb barrier due to the presence of the distorted waves in the initial and final channels. Usually to analyze such reactions the distorted-wave-Born-approximation (DWBA) is used. The DWBA amplitude for peripheral reactions is parametrized in terms of the asymptotic normalization coefficient (ANC) of the bound state $B= (n\,A)$. In this paper, I prove that the sub-Coulomb $A(d,p)B$ reaction amplitude, which is a solution of the three-body Faddeev equations in the Alt-Grassberger-Sandhas (AGS) form, is peripheral if a peripheral is the corresponding DWBA amplitude. Hence the Faddeev's reaction amplitude for the sub-Coulomb $A(d,p)B$ reactions can also be parametrized in terms of the ANC of the $(n\,A)$ bound state. First, I consider the original AGS equations with separable potentials and prove that such equations are peripheral at sub-Coulomb energies. After that, the two-particle  AGS equations are derived for the general potentials for sub-Coulomb transfer reactions. The effective AGS potentials are expressed in terms of the DWBA amplitudes for the sub-Coulomb reactions. Again, I demonstrate that the amplitude of the $A(d,p)B$ transfer reaction obtained from the AGS equation is peripheral and can be parametrized in terms of the ANC for the $(nA)$ bound state because the corresponding DWBA amplitude is peripheral. Finally, the AGS equations are generalized by including the optical nuclear potentials in the same manner as it is done in the DWBA. The obtained two-particle AGS equations contain the DWBA effective potentials with distorted waves generated by the sum of the nuclear optical and the channel Coulomb potentials. The AGS equation for the $A(d,p)B$ reactions is analyzed above the Coulomb barrier and it is shown again that the reaction amplitude satisfying generalized AGS equation with optical potentials depends on the ANC if a peripheral is the DWBA amplitude. 
The two-body AGS equations are generalized by including the intermediate three-body continuum and more than one bound state in each channel. 
\end{abstract}

\maketitle

\date{Today}

\section{Introduction}
\label{Standardapproach1}

Let us consider the transfer reaction in the three-body model of three non-identical constituent structureless particles:
\begin{align}
\alpha + (\beta\,\gamma)  \to \beta + (\alpha\,\gamma),
\label{reaction1}
\end{align}
where $(\beta\,\gamma)$ is the bound state of particles $\beta$ and $\gamma$. 
The general expression for the reaction amplitude in the center-off-mass (c.m.) of  reaction (\ref{reaction1}) in the three-body model is
\begin{align}
{\cal T}_{\beta\,\alpha}({\rm {\bf q}}_{\beta},\,{\rm {\bf q}}_{\alpha};z) = <\psi^{(0)}_{{\rm {\bf q}}_{\beta}}\,\varphi_{\beta}^{(R)}\big|U_{\beta\,\alpha}^{(R)(+)}(z)\big|\varphi_{\alpha}^{(R)}\,\psi_{{\rm {\bf q}}_{\alpha}}^{(0)}>, 
\label{reactampl1}
\end{align}

\begin{align}
U_{\beta\,\alpha}^{(R)(+)}(z) ={\overline V}_{\beta}^{(R)} +
{\overline V}_{\beta}^{(R)}\,G^{(R)}(z)\,{\overline V}_{\alpha}^{(R)}
\label{Uplus1}
\end{align}
is the transition operator,
$V^{(R)}= \sum\limits_{\nu=\alpha,\,\beta\,\gamma}\,V_{\nu}^{(R)}$,
\begin{align}
G^{(R)}(z)= \frac{1}{z - K - V^{(R)}}
\label{Greenfunction1}
\end{align}
is the three-body Green function resolvent, $E$ and $K$ are the total energy and kinetic energy operator of the three-body system. 
The superscript $R$ means that I use the screened Coulomb potentials. Correspondingly, all other functions, which depend on the screened Coulomb potentials, also have the superscript $R$. I use the following supplemental notation usually accepted in few-body papers: for a one-body quantity an index $\alpha$ characterizes the particle
$\alpha$, for a two-body quantity the pair of particles $(\beta + \gamma)$, with $\beta,\,\gamma \not = \alpha$ and finally for a three-body 
quantity the two-fragment partition $\alpha + (\beta\,\gamma)$ describing free particle $\alpha$ and the
bound state $(\beta\,\gamma)$. 
$\psi_{{\rm {\bf q}}_{\alpha}}^{(0)}$ is the plane wave describing the relative motion of particles $\alpha$ and the bound state $(\beta\,\gamma)$ of pair $\alpha$ with  the relative momentum ${\rm {\bf q}}_{\alpha}$,  $\,\varphi_{\alpha}^{(R)}$ is the bound state of particles of the pair $\alpha$.
\begin{align}
&{\overline V}_{\alpha}^{(R)}  =  V^{(R)} - V_{\alpha}^{(R)},           \nonumber\\ 
&V_{\alpha}^{(R)} \equiv V_{\beta\,\gamma}^{(R)} = V_{\alpha}^{N} + V_{\alpha}^{C(R)},
\label{Vpot1}
\end{align}
where $V_{\alpha}^{N} \equiv V_{\beta\,\gamma}^{N}$ and $\,V^{C(R)}_{\alpha} \equiv V_{\beta\,\gamma}^{C(R)}$ are the nuclear and screened Coulomb interaction potentials of particles of pair $\alpha$. 
Note that the plane waves in Eq. (\ref{reactampl1}) appear only for the screened Coulomb potentials.
  
Taking into account that  
\begin{align}
\Psi_{\alpha}^{(R)(+)}= \big(1 + G^{(R)}(E)\,{\overline V}_{\alpha}^{(R)} \big)\,\varphi_{\alpha}^{(R)}\,\psi_{{\rm {\bf q}}_{\alpha}}^{(0)}
\label{Psialpa1}
\end{align}
one gets for the reaction amplitude
\begin{align}
{\cal T}_{\beta\,\alpha}({\rm {\bf q}}_{\beta},\,{\rm {\bf q}}_{\alpha};E)= <\psi_{{\rm {\bf q}}_{\beta}}^{(0)}\,\varphi_{\beta}^{(R)}\big|{\overline V}_{\beta}^{(R)}\big|\Psi_{\alpha}^{(R)(+)}>.
\label{reactampl2}
\end{align}

Thus to calculate ${\cal T}_{\beta\,\alpha}$ one needs to find the exact scattering wave function $\Psi_{\alpha}^{(R)(+)}$,
which is a solution of the equation  
\begin{align}
\Psi_{\alpha}^{(R)(+)} = \varphi_{\alpha}^{(R)}\,\psi_{\alpha}^{(0)}   
 + G_{\alpha}^{(R)}(E)\,{\overline V}_{\alpha}^{(R)}\,\Psi_{\alpha}^{(R)(+)},
\label{Psialpha2}
\end{align}
where 
\begin{align}
G_{\alpha}^{(R)}(z)= \frac{1}{z - K - V_{\alpha}^{(R)}}.
\label{Greenfunction1}
\end{align}

This equation does not have a unique solution because one can add to $\Psi_{\alpha}^{(R)(+)}$ a linear combination of solutions of the homogeneous equations
\begin{align}
\Psi_{\nu}^{(R)(+)} =   
 G_{\alpha}^{(R)}(E)\,{\overline V}_{\alpha}^{(R)}\,\Psi_{\nu}^{(R)(+)}, \quad \nu = \beta,\,\gamma \not= \alpha.
\label{Psinu1}
\end{align} 

The way to find the transfer reaction amplitude unambiguously was suggested by Faddeev \cite{Faddeev} by using the coupled Faddeev integro-differential equations in the three-body problem. This seminal work by Faddeev showed how to solve exactly the three-body quantum-mechanical problem and opened a new field in physics: few-body physics. In his original work Faddeev considered $3\,{\it particles} \to 3\,{\it particles}$ case.
Later on in \cite{alt1967} Alt, Grassberger and Sandhas modified the Faddeev equations by transforming them into equations describing $2\,{\it particles} \to 2\,{\it particles}$ processes. These modified equations are called the Faddeev equations in the AGS form. An important advantage of the AGS formalism is that it reduces the three-body Faddeev equations to the two-particle form
when the separable potentials are used. In \cite{alt78} the AGS equations were modified by including the Coulomb interaction for the processes involving two charged particles and a neutron. 

In this paper, I use the AGS formalism for the analysis of the $A(d,p)B$ peripheral reactions, which
allows one to extract the ANC $C_{nA}$ of the bound state $B=(n\,A)$ of the final nucleus $B$\cite{reviewpaper}. Moreover, the deuteron stripping reactions on unstable nuclei $A$ in the inverse kinematics provide a unique tool to obtain the spectroscopic information about the $(n\,A)$ bound states and resonances. Usually, for the analysis of such reactions the traditional DWBA, adiabatic distorted wave (ADWA) \cite{Johnson} or its extensions, continuum-discretized-coupled-channel (CDCC) method \cite{Austern}, are used. 

First, I analyze the sub-Coulomb $A(d,p)B$ reactions which are peripheral due to the Coulomb barriers in the initial and final states. Both non-local separable and local general potentials are considered. It is shown that the AGS amplitude for the sub-Coulomb $A(d,p)B$ reaction is peripheral if  the effective potential given by the DWBA amplitude is peripheral. 
Then, for the first time, the AGS equations are modified by inserting the optical potentials as it is done in the DWBA. The effective potentials in new generalized AGS equations are expressed in terms of the DWBA amplitudes. 

The system of units in which $\hbar=c=1$ is used throughout the paper.\\

\section{AGS equations with separable potentials}
\label{AGSseparable1}

Let us consider the system of three distinguishable constituent particles, $1,\,2,\,3$ with masses $m_{\nu},\,\nu=1,2,3$. Moreover, we assume that particles $1$ and $2$ are charged with charges $Z_{1}e$ and $Z_{2}e$ satisfying $Z_{1}\,Z_{2} >0$.  In this case only one Coulomb potential $V_{3}^{C} \equiv V_{12}^{C}$ enters the AGS equations. In what follows, I use the following notations: for a one-body quantity an index $\alpha$ characterizes the particle
$\alpha$, for a two-body quantity the pair of particles $(\beta + \gamma)$, with $\beta,\,\gamma \not = \alpha$ and finally for a three-body 
quantity the two-fragment partition $\alpha + (\beta\,\gamma)$ describing free particles $\alpha$ and the bound state $(\beta\,\gamma)$. 

I assume here, for simplicity, that the nuclear interaction potential between the particles of the pair $\alpha$ is given by the rank one separable potential:
\begin{align}
V_{\alpha}^{N}= |g_{\alpha}>\,\lambda_{\alpha}\,<g_{\alpha}|, \qquad  \alpha \not = 3,
\label{separpot1}
\end{align}
\begin{align}
V_{3}= |g_{3}\,\lambda_{3}\,<g_{3}| V_{3}^{C(R)},
\label{separpotC31}
\end{align}
where $g_{\alpha}$ is the form factor of the pair $\alpha$ and $\lambda_{\alpha}$ is the strength parameter,
$V_{3}^{C(R)}$ is the screened Coulomb interaction potential between particles $1$ and $2$. 
Extension for the arbitrary separable rank potential is straightforward \cite{alt2002}.

Then the Faddeev equations for the transition operators take the form
\begin{align}
U_{\beta\,\alpha}^{(R)(+)}(z) = {\cal V}_{\beta\,\alpha}                       
+ \sum\limits_{\nu}\,{\cal V}_{\beta\,\nu}\,S_{\nu}(z){\overline G}_{\nu}(z + \varepsilon_{\nu})\,U_{\nu\,\alpha}^{(R)(+)}(z),
\label{Ubaeq1}
\end{align} 
where 
\begin{align}
{\overline G}_{\nu}(z_{\nu}) = \frac{1}{z + \varepsilon_{\nu} - {\overline K}_{\nu}}.
\label{overlKnu1}
\end{align}

Taking into account the matrix elements from both sides of Eq. (\ref{Ubaeq1}) and the definition (\ref{reactampl1}) of the reaction amplitude we get the two-particle AGS equations 
\begin{widetext}
\begin{align}
&{\cal T}_{\beta\,\alpha}({\rm {\bf q}}_{\beta},\,{\rm {\bf q}}_{\alpha};z) = 
{\cal V}_{\beta\,\alpha}({\rm {\bf q}}_{\beta},\,{\rm {\bf q}}_{\alpha};z)            
+ \sum\limits_{\nu}\,\int \frac{{\rm d}{\rm {\bf p}}_{\nu}}{(2\,{\pi})^{3}}{\cal V}_{\beta\,\nu}({\rm {\bf q}}_{\beta},{\rm {\bf p}}_{\nu};z)\,\frac{S_{\alpha}(z - \frac{p_{\alpha}^{2}}{2\,{\cal M}_{\alpha}})}{z +\varepsilon_{\nu} - p_{\nu}^{2}/(2\,{\cal M}_{\nu})}\,{\cal T}_{\nu\,\alpha}({{\rm {\bf p}}_{\nu},{\rm {\bf q}}_{\alpha};z}),
\label{TbaInit1}
\end{align}
\end{widetext}
where $\,{\rm {\bf q}}_{\alpha}\,$ is on-the-energy-shell (ONES) relative momentum in the channel $\,\alpha$, that is, the relative momentum of particle $\,\alpha\,$ and the bound state $\,(\beta\,\gamma)$, which is related to the ONES energy of the three-body system as
$E= \frac{q_{\alpha}^{2}}{2\,{\cal M}_{\alpha}} - \varepsilon_{\alpha}$, $\varepsilon_{\alpha}$ is the binding energy of the bound state $(\beta\,\gamma),\,$ $\,{\cal M}_{\alpha} =  m_{\alpha}\,m_{\beta\,\gamma}/{\cal M}$ is the reduced mass of the particles in the channel $\,\alpha$, $\,m_{\beta\,\gamma}= m_{\beta} + m_{\gamma}$, $\,{\cal M}= \sum\limits_{\alpha=1}^{3}\,m_{\alpha}$ is the total mass of the three-body system.

To obtain the ONES AGS equations from Eqs (\ref{TbaInit1}) one needs to take the limit $ z \to E_{+}= E + i0$. 
Also
\begin{align}
&S_{\alpha}^{-1}(z_{\alpha}) = \big[<g_{\alpha} \big|{\hat G}_{0}({-\varepsilon}_{\alpha})\,
{\hat G}_{0}(z_{\alpha}) \big| g_{\alpha}>\big]                   \nonumber\\
&= -\int \frac{ {\rm d}{\rm {\bf k}}_{\alpha} }{(2\,\pi)^{3}}\,
\frac{g_{\alpha}^{2}({\rm {\bf k}}_{\alpha})  }{ (\frac{k_{\alpha}^{2}}{2\,\mu_{\alpha}} + \varepsilon_{\alpha})(z_{\alpha} - \frac{k_{\alpha}^{2}}{2\,{\mu}_{\alpha}})},
\label{Salpha1}
\end{align} 
where $\mu_{\alpha} = m_{\beta}\,m_{\gamma}/m_{\beta\,\gamma}$.

\begin{align}
&S_{3}(z_{3})^{-1} = \big[<g_{3} \big|{\hat G}^{C(R)}({-\varepsilon}_{3})\,
{\hat G}^{C(R)}(z_{3}) \big| g_{3}>\big],
\label{SR31} 
\end{align}
\begin{align}
{\hat G}_{0}(z_{\alpha})=\frac{1}{z_{\alpha} - K_{\alpha}},
\label{hatG01}
\end{align}
\begin{align}
{\hat G}^{C(R)}(z_{\alpha})=\frac{1}{z_{\alpha} - K_{\alpha} - V_{\alpha}^{C(R)}}.
\label{hatG01}
\end{align} 
For example, $\,z_{\alpha}= z- \frac{p_{\alpha}^{2}}{2\,{\cal M}_{\alpha}}$ for $\,\,S_{\alpha}\big(z- \frac{p_{\alpha}^{2} }{2\,{\cal M}_{\alpha}}\big)$.
$K_{\alpha}$ is the kinetic energy operator of the relative motion of particles $\beta$ and $\gamma$.
For our purposes it is important that $S_{\alpha}(-\varepsilon_{\alpha})=1$. 

Half-off-energy-shell (HOES) effective potentials ${\cal V}_{\beta\,\alpha}({\rm {\bf q}}_{\beta},{\rm {\bf p}}_{\alpha};E)$ in the AGS equations are \cite{alt1980}:
\begin{widetext}
\begin{align}
&{\cal V}_{\beta\,\alpha}({\rm {\bf q}}_{\beta},{\rm {\bf p}}_{\alpha};E)=  <{\rm {\bf q}}_{\beta}|<g_{\beta}|\,{\overline \delta}_{\beta\,\alpha}\,\big[ G_{0}(z) + \big(\delta_{\beta\,3} + \delta_{\alpha\,3}  \big) G_{0}(E)\,T_{3}^{C(R)}(z)\,G_{0}(E)\big] \nonumber\\
&+ \delta_{\beta\,\alpha}\,{\overline \delta}_{\beta\,3}G_{0}(E)\,T_{3}^{C(R)}(E)\,G_{0}(E)\,
+{\overline \delta}_{\beta\,\alpha}\,{\overline \delta}_{\beta\,3}\,{\overline \delta}_{\alpha\,3}  G_{0}(E)\,T_{3}^{C(R)}(E)\,G_{0}(E)
|g_{\alpha}>|{\rm {\bf  p}}_{\alpha}>,
\label{calV1}
\end{align}
\end{widetext} 
\begin{align}
G_{0}(z) = \frac{1}{z - K}.
\label{G01}
\end{align}
By taking ${\rm {\bf p}}_{\alpha}= {\rm {\bf q}}_{\alpha}$ in Eq. (\ref{calV1}) one gets ONES ${\cal V}_{\beta\,\alpha}({\rm {\bf q}}_{\beta},\,{\rm {\bf q}}_{\alpha};z)$, which is the first term on the right-hand-side of Eq. (\ref{TbaInit1}).

The main advantage of the AGS Eqs (\ref{TbaInit1}) is that they reduce the three-body Faddeev equations to the  two-body ones. This us achieved by using the separable potentials what allows one to single out explicitly the bound-state poles.  

However, Eqs (\ref{TbaInit1}) have too strong Coulomb singularity in the elastic scattering part. The Coulomb  elastic scattering  effective potential $ <{\rm {\bf q}}_{\beta}|<g_{\beta}| \delta_{\beta\,\alpha}\,{\overline \delta}_{\beta\,3}G_{0}(z)\,T_{3}^{C(R)}(z)\,G_{0}(z)|g_{\alpha}>|{\rm {\bf  p}}_{\alpha}>$ 
has a strong singularity $1/q^{2}$ in the transfer momentum plane \cite{alt78}. Coincidence of this singularity with the singularity of the Green function $G_{0}$ generates a non-compact singularity. 
To remove this singularity in \cite{alt78} the two-potential equation was applied what leads to the  AGS equations for the short-ranged Coulomb-modified reaction amplitude with separable potentials (\ref{separpot1}) and (\ref{separpotC31}). 

To this end the new transition operator is introduced in which the Coulomb channel potential $U_{\nu}^{C(R)}$ is subtracted from each ${\overline V}_{\nu}^{(R)}$: 
\begin{align}
U_{\beta\,\alpha}^{NC(R)(+)}(z) = \Delta{\overline V}_{\beta}^{(R)} +
\Delta{\overline V}_{\beta}^{(R)}\,G^{(R)}(z)\,\Delta{\overline V}_{\alpha}^{(R)},
\label{tildeUCRplus1}
\end{align}
\begin{align}
{\Delta{\overline V}_{\alpha}^{(R)}} = {\overline V}_{\alpha} - U_{\alpha}^{C(R)} 
={\overline V}_{\alpha}^{N} + \Delta {\overline V}_{\alpha}^{C(R)},
\label{DeltaunderValpha1}
\end{align}
\begin{align}
{\Delta {{\overline V}}^{C(R)}_{\alpha}} = {\overline V}_{\alpha}^{C(R)} - U_{\alpha}^{C(R)}.
\label{Ua11}
\end{align}
Here, $\,{\overline V}_{\alpha}$ is determined in Eq. (\ref{Vpot1}), $\,{\overline V}_{\alpha}^{N}
= V_{\beta}^{N} + V_{\gamma}^{N}$, 
$\,{\overline V}_{\alpha}^{C(R)} = V_{\beta}^{C(R)} + V_{\gamma}^{C(R)}$, $\,U_{\alpha}^{C(R)}$ is the screened  Coulomb $\alpha$-channel potential acting between the particle $\,\alpha$ and the c.m. of the bound state $(\beta\,\gamma)$. I remind that the Greek indices $\alpha,\,\beta,\gamma$ can be $1,\,2,\,3$.\
The superscripts $N$ and $C$ mean nuclear and Coulomb, correspondingly.

The original transition operator $U_{\beta\,\alpha}^{(+)}(z)$ is related to $U_{\beta\,\alpha}^{NC(R)(+)}(z)$ 
as
\begin{widetext}
\begin{align}
U_{\beta\,\alpha}^{(+)}(z) = \delta_{\beta\,\alpha}\,T_{\alpha}^{C(R)}(z) + \big(1+ T_{\beta}^{C(R)}(z)\,G_{0}(z)\big)
\,U_{\beta\,\alpha}^{NC(R)(+)}(z)\,\big(1+ G_{0}(z)\,T_{\alpha}^{C(R)}(z)\big),
\label{UbaUncba1}
\end{align}
\end{widetext}
where $\,T_{\alpha}^{C(R)}(z)\,$ is the $\,\alpha + (\beta\,\gamma)\,$ Coulomb elastic scattering $\,T\,$-matrix operator generated by the screened Coulomb channel potential $\,U_{\alpha}^{C(R)}$. 

Then the Faddeev equations (\ref{Ubaeq1}) for $U_{\beta\,\alpha}^{(+)}(z)$ are transformed into the equations for  
the transition operator $U_{\beta\,\alpha}^{NC(R)(+)}(z)$:
\begin{align}
&U_{\beta\,\alpha}^{NC(R)(+)}(z) = {\cal V}^{NC(R)}_{\beta\,\alpha}(z)                        \nonumber\\
&+ \sum\limits_{\nu}\,{\cal V}^{NC(R)}_{\beta\,\nu}(z)\,S_{\nu}(z)\,{\overline G}_{\nu}^{C(R)}(z + \varepsilon_{\nu})\,U_{\nu\,\alpha}^{NC(R)(+)}(z),
\label{UbaNCR1}
\end{align} 
where 
\begin{align}
{\overline G}_{\nu}^{C(R)}(z + \varepsilon_{\nu}) = \frac{1}{z  + \varepsilon_{\nu} - {\overline K}_{\nu} - U_{\nu}^{C(R)}}.
\label{overlineKnuCR1}
\end{align}

Taking the matrix elements from both sides of Eq. (\ref{UbaUncba1}) we get the two-particle AGS  integral equations for the short-range reaction amplitudes
\begin{widetext}
\begin{align}
&{\cal T}_{\beta\,\alpha}^{NC(R)}({\rm {\bf q}}_{\beta}^{C(R)(-)},{\rm {\bf q}}_{\alpha}^{C(R)(+)};\,z) 
=  <\psi_{{\rm {\bf q}}_{\beta}}^{C(R)(-)}\,\varphi_{\beta}^{(R)}\big|U_{\beta\,\alpha}^{NC(R)(+)}(z)\big| \varphi_{\alpha}^{(R)}\,\psi_{{\rm {\bf q}}_{\alpha}}^{C(R)(+}>                        \nonumber\\
&= {\cal V}_{\beta\,\alpha}^{C(R)}({\rm {\bf q}}_{\beta}^{C(R)(-)},{\rm {\bf q}}_{\alpha}^{C(R)(+)};z)                                                           \nonumber\\
& + \sum\limits_{\nu}\,\int\,\frac{{\rm d}{\rm {\bf p}}_{\nu}}{(2\,{\pi})^{3}}{\cal {\tilde V}}_{\beta\,\nu}^{NC(R)}({\rm {\bf q}}_{\beta}^{C(R)(-)},{\rm {\bf p}}_{\nu}^{C(R)(-)};z)\,\frac{1}{z +\varepsilon_{\nu} - p_{\nu}^{2}/(2\,{\cal M}_{\nu})}\,{\cal T}_{\nu\,\alpha}^{NC(R)}({{\rm {\bf p}}_{\nu}^{C(R)(-)},{\rm {\bf q}}_{\alpha}^{C(R)(+)};z}).
\label{AGCCoulm1}
\end{align} 
\end{widetext}

The short-range reaction amplitude $\,{\cal T}_{\beta\,\alpha}^{NC(R)}({\rm {\bf q}}_{\beta}^{C(R)(-)},{\rm {\bf q}}_{\alpha}^{C(R)(+)};\,z)\,$ with the ONES initial and final momenta is defined as
\begin{align}
&{\cal T}_{\beta\,\alpha}^{NC(R)}({\rm {\bf q}}_{\beta}^{C(R)(-)},{\rm {\bf q}}_{\alpha}^{C(R)(+)};\,z)
\nonumber\\
&= <\psi_{{\rm {\bf p}}_{\beta}}^{C(R)(-)}\,\varphi_{\beta}^{(R)}\big|U_{\beta\,\alpha}^{NC(R)(+)}(z)\big| \varphi_{\alpha}^{(R)}\,\psi_{{\rm {\bf q}}_{\alpha}}^{C(R)(+}>.
\label{calTbetaalpha1}
\end{align}
The subtraction of the channel potential $U_{\nu}^{C(R)}$ from ${\overline V}_{\nu}^{(R)}$ leads to the appearance of the Coulomb distorted waves in the bra and ket states:
$\psi_{{\rm {\bf q}}_{\alpha}}^{C(R)(+)} = {\rm {\bf q}}_{\alpha}^{C(R)(+)}$ is the Coulomb scattering wave function of particle $\alpha$ and the bound state $\,(\beta\,\gamma)$,$\,\,{\rm {\bf p}}_{\alpha}$ is the off-the-energy-shell (OFES) relative momentum of the particles in the channel $\alpha$. 

Note that the amplitudes  ${\cal T}_{\nu\,\alpha}^{NC(R)}({{\rm {\bf p}}_{\nu}^{C(R)(-)},{\rm {\bf q}}_{\alpha}^{C(R)(+)};z})$ under the integral sign are HOES.
I also assume that only one bound state can be populated in each pair of the particles.
An additional number of the bound states can be added in a straightforward manner
\cite{alt1980,muk2012}, see also section  \ref{generalizedAGSCDCC1}.

The effective potentials in Eqs. (\ref{AGCCoulm1}) on the right-hand-side under the integral sign are given by
\cite{alt1980}
\begin{widetext}
\begin{align}
&{\cal {\tilde V}}_{\beta\,\alpha}^{NC(R)}({\rm {\bf q}}_{\beta}^{C(R)(-)},{\rm {\bf p}}_{\nu}^{C(R)(-)};z) = <\psi_{{\rm {\bf q}}_{\beta}}^{C(R)(-)}({\rm {\bf p}}_{\beta}')\big|{\cal {\tilde V}}_{\beta\,\alpha}^{NC(R)}({\rm {\bf p}}_{\beta}',{\rm {\bf p}}_{\alpha}';z)\,\big|\psi_{{\rm {\bf p}}_{\alpha}}^{C(R)(-)}({\rm {\bf p}}_{\alpha}')>,
\label{Vbatr1}
\end{align} 
\end{widetext}
where
\begin{widetext}
\begin{align}
&{\cal {\tilde V}}_{\beta\,\alpha}^{NC(R)}({\rm {\bf p}}_{\beta}',{\rm {\bf p}}_{\alpha}';z)= \big\{ <{\rm {\bf p}}_{\beta}'|<g_{\beta}|\,{\overline \delta}_{\beta\,\alpha}\,\big[ G_{0}(z) + \big(\delta_{\beta\,3} + \delta_{\alpha\,3}  \big) G_{0}(z)\,T_{3}^{C(R)}(z)\,G_{0}(z)\big] \nonumber\\
&+ \delta_{\beta\,\alpha}\,{\overline \delta}_{\beta\,3}G_{0}(z)\,T_{3}^{C(R)}(z)\,G_{0}(z)\,
+{\overline \delta}_{\beta\,\alpha}\,{\overline \delta}_{\beta\,3}\,{\overline \delta}_{\alpha\,3}  G_{0}(z)\,T_{3}^{C(R)}(z)\,G_{0}(z)
|g_{\alpha}>|{\rm {\bf  p}}_{\alpha}'>\big\}\,S_{\alpha}\big(z- \frac{ {p_{\alpha}'}^{2} }{2\,{\cal M}_{\alpha}}\big)                                                                                   \nonumber\\
&-\delta_{\beta\,\alpha}\,{\overline \delta}_{\beta\,3}\,<{\rm {\bf p}}_{\beta}'|<g_{\beta}|\,\big[ G_{0}(z)\,U_{\alpha}^{C(R)}\big) \,G_{0}(z)\,
|g_{\alpha}>|{\rm {\bf  p}}_{\alpha}'>.
\label{calV1}
\end{align}
\end{widetext}
Here, $T_{3}^{C(R)}(z) \equiv T_{12}^{C(R)}(z)$ is the $T$-operator of the Coulomb elastic scattering of the particles $2$ and $3$ interacting via the screened Coulomb potential $V_{3}^{C(R)}$,

ONES $S_{\alpha}\big(E_{+} - \frac{q_{\alpha}^{2} }{2\,{\cal M}_{\alpha}}\big) =
S_{\alpha}\big(- \varepsilon_{\alpha}  \big)=1$ and 
\begin{widetext} 
\begin{align}
&{\cal {\tilde V}}_{\beta\,\alpha}^{NC(R)}({\rm {\bf q}}_{\beta},{\rm {\bf q}}_{\alpha};E_{+})= {\cal V}_{\beta\,\alpha}^{NC(R)}({\rm {\bf q}}_{\beta},{\rm {\bf q}}_{\alpha};E_{+})    \nonumber\\
&=<{\rm {\bf q}}_{\beta}|<g_{\beta}|\,{\overline \delta}_{\beta\,\alpha}\,\big[ G_{0}(E_{+}) + \big(\delta_{\beta\,3} + \delta_{\alpha\,3}  \big) G_{0}(E_{+})\,T_{3}^{C(R)}(E_{+})\,G_{0}(E_{+})\big]|g_{\alpha}>|{\rm {\bf  q}}_{\alpha}>  \nonumber\\
&+ <{\rm {\bf q}}_{\beta}|<g_{\beta}| \delta_{\beta\,\alpha}\,{\overline \delta}_{\beta\,3}G_{0}(E_{+})\,\big(T_{3}^{C(R)}(E_{+}) - U_{\alpha}^{C(R)}\big) \,G_{0}(E_{+})\,|g_{\alpha}>|{\rm {\bf  q}}_{\alpha}>                                                             \nonumber\\
&+ <{\rm {\bf q}}_{\beta}|<g_{\beta}|{\overline \delta}_{\beta\,\alpha}\,{\overline \delta}_{\beta\,3}\,{\overline \delta}_{\alpha\,3}  G_{0}(E_{+})\,T_{3}^{C(R)}(E_{+})\,G_{0}(E_{+})|g_{\alpha}>|{\rm {\bf  q}}_{\alpha}>.       \nonumber\\
\label{Vbaonsh1}
\end{align}
\end{widetext}
Taking into account that ($\alpha \not=3$)
\begin{align}
G_{0}(E_{+})|g_{\alpha}|{\rm {\bf q}}_{\alpha}>= - |{\rm {\bf q}}_{\alpha}> \frac{1}{K_{\alpha} + \varepsilon_{\alpha}}\,|g_{\alpha}>= |\varphi_{\alpha}>|{\rm {\bf q}}_{\alpha}>,
\label{varphichi1}
\end{align}
where $\varphi_{\alpha}$ is the bound-state wave function of the bound state $(\beta\,\gamma)$,
one can rewrite the ONES effective potential as
\begin{widetext} 
\begin{align}
&{\cal V}_{\beta\,\alpha}^{NC(R)}({\rm {\bf q}}_{\beta},{\rm {\bf q}}_{\alpha};E_{+})= {\overline \delta}_{\beta\,\alpha}\,
\big[<{\rm {\bf q}}_{\beta}|<\varphi_{\beta}|g_{\alpha}>|{\rm {\bf q}}_{\alpha}>  +  \big(\delta_{\beta\,3} + \delta_{\alpha\,3}\big)\,
<{\rm {\bf q}}_{\beta}|<\varphi_{\beta}|\, T_{3}^{C(R)}(E_{+})\,|\varphi_{\alpha}>|{\rm {\bf  q}}_{\alpha}> \big]                                         \nonumber\\
&+ \delta_{\beta\,\alpha}\,{\overline \delta}_{\beta\,3}\,\big[<{\rm {\bf q}}_{\beta}|<\varphi_{\beta}| \big(T_{3}^{C(R)}(E_{+}) - U_{\alpha}^{C(R)}\big) \,|\varphi_{\alpha}>|{\rm {\bf  q}}_{\alpha}> \big]                                                            \nonumber\\
&+ {\overline \delta}_{\beta\,\alpha}\,{\overline \delta}_{\beta\,3}\,{\overline \delta}_{\alpha\,3}\,\big[<{\rm {\bf q}}_{\beta}|<\varphi_{\beta}|T_{3}^{(CR)}(E_{+})\,|\varphi_{\alpha}>|{\rm {\bf  q}}_{\alpha}> \big].     
\label{Vbaonsh11}
\end{align}
\end{widetext}
The subtraction of $U_{\alpha}^{C(R)}$ from $T_{3}^{C(R)}$ compensates the most singular term in the latter \cite{alt1980}.  Also $<{\rm {\bf q}}_{\beta}|<\varphi_{\beta}|g_{\alpha}>|{\rm {\bf q}}_{\alpha}>= <{\rm {\bf q}}_{\beta}|<g_{\beta}|\varphi_{\alpha}>|{\rm {\bf q}}_{\alpha}>$.

\begin{figure}[tbp] 
\includegraphics[width=4.0in,height=4.5in]{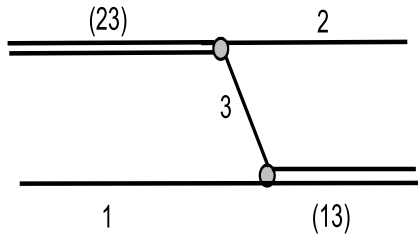}
  \caption{Pole diagram describing the $(23) + 1 \to 2 + (13)$ transfer reaction. }
\label{fig_pole(23)2}
\end{figure}

\begin{figure}[tbp] 
\includegraphics[width=4in,height=4.5in]{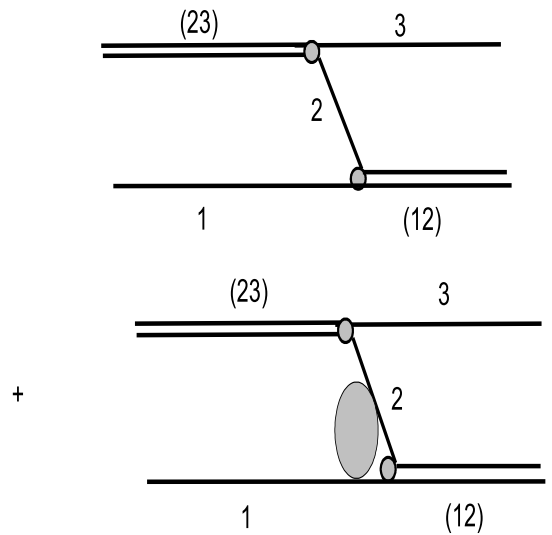}
  \caption{Sum of the diagrams describing the $(23) + 1 \to 3+ (12)$ reaction. The dashed bulb is the $1+2$ Coulomb scattering amplitude.}
\label{fig_pole(23)3}
\end{figure}

\begin{figure}[tp] 
  \centering
\includegraphics[width=4in,height=4.5in]{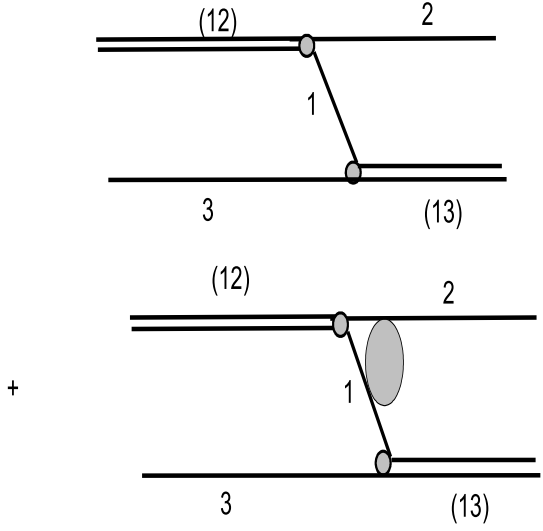}
  \caption{Sum of the diagrams describing the transfer reaction $(12) + 3 \to 2+ (13)$. The dashed bulb is the $1 +2$ Coulomb scattering amplitude. }
\label{fig_(12)2Coul}
\end{figure}

I remind the reader that particle $3$ is assumed to be the neutron.
Let us consider the first bracket in Eq. (\ref{Vbaonsh11}). 
For $\alpha,\,\beta \not=3$ the term $<{\rm {\bf q}}_{\beta}|<\varphi_{\beta}|g_{\alpha}>|{\rm {\bf q}}_{\alpha}>$ is described by the pole diagram in Fig.  \ref{fig_pole(23)2}.  
For $\beta=3$ the first bracket is described by the sum of the diagrams in Fig. \ref{fig_pole(23)3}.
For $\alpha=3$ the first bracket is described by the sum of the diagrams in 
Fig. \ref{fig_(12)2Coul}. 

One can introduce the Coulomb-modified vertex form factor of the pair $\nu=3$, which takes into account the Coulomb interaction between the particles $1$ and $2$: 
\begin{align}
|\phi_{3}(z_{3})> = \big[ 1 + T_{3}^{C(R)}(z_{3})\,{\hat G}_{0}(z_{3}) \big] \big|g_{3}>,
\label{phinu31}
\end{align}
\begin{align}
&\big|{\rm {\bf q}}_{\alpha}>\big|\phi_{3}(z- \frac{q_{\alpha}^{2}}{2\,{\cal M}_{\alpha}})>   \nonumber\\
&= \big[ 1 + T_{3}^{C(R)}(z)\,G_{0}(z) \big] \big|g_{3}(z)>\big|{\rm {\bf q}}_{\alpha}>.
\label{phiq1}
\end{align}
The properties of $|\phi_{3}(z_{3})>$ were discussed in details in \cite{muk2000}.

Then for $\alpha=3$ the first bracket in Eq. (\ref{Vbaonsh11}) can be rewritten as
\begin{widetext}
\begin{align}
<{\rm {\bf q}}_{\beta}|<\varphi_{\beta}|g_{3}>|{\rm {\bf q}}_{3}>  +  
<{\rm {\bf q}}_{\beta}|<\varphi_{\beta}|\, T_{3}^{C(R)}(E_{+})\,|\varphi_{3}>|{\rm {\bf  q}}_{3}>  
=<\varphi_{\beta}({\rm {\bf k}}_{\beta}; -\varepsilon_{\beta})|\phi_{\alpha}({\rm {\bf k}}_{\alpha}; -\varepsilon_{\alpha})>, 
\label{alpha31}
\end{align}
\end{widetext}
where $\varphi_{\beta}({\rm {\bf k}}_{\beta}; -\varepsilon_{\beta})$ is the Fourier transform of the 
bound-state wave function of the pair $\beta$ (bound state $(\alpha\,\gamma)$) with the binding energy $\varepsilon_{\beta}$; $\phi_{\alpha}({\rm {\bf k}}_{\alpha}; -\varepsilon_{\alpha})$ is the Fourier transform of the Coulomb-modified form factor of the pair $\alpha=3$  with the binding energy $\varepsilon_{\alpha}$,
$\,{\rm {\bf k}}_{\alpha} = {\rm {\bf q}}_{\beta}  + \frac{m_{\beta}}{m_{\beta\,\gamma}}\,{\rm {\bf q}}_{\alpha}$ is the relative momentum of particles $\beta$ and $\gamma$ in the vertex $(\beta\,\gamma) \to \beta + \gamma,\,\,$ $\,{\rm {\bf k}}_{\beta} = {\rm {\bf q}}_{\alpha}  + \frac{m_{\alpha}}{m_{\alpha\,\gamma}}\,{\rm {\bf q}}_{\beta}$ is the relative momentum of particles 
$\alpha$ and $\gamma$.
Since ${\rm {\bf q}}_{\alpha}$ is the relative momentum of particle $\alpha$ and the pair $(\beta\,\gamma)$, in the c.m. ${\rm{\bf q}}_{\alpha}$ is the momentum of the particle $\alpha$ and $-{\rm {\bf q}}_{\alpha}$ is the momentum of the pair $(\beta\,\gamma)$.

\begin{figure}[tbp] 
\includegraphics[width=4.0in,height=4.5in]{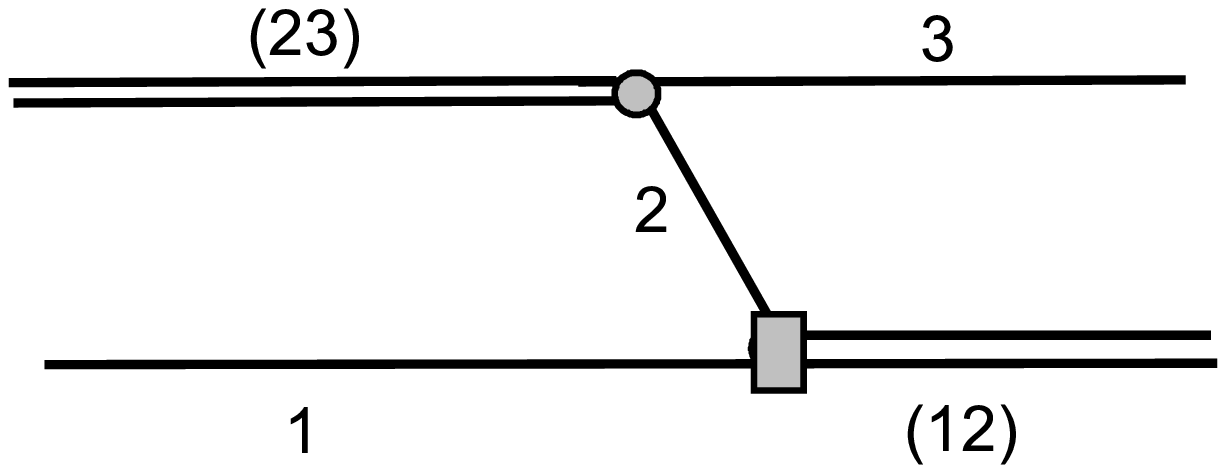}
  \caption{Diagram describing the $(23) + 1 \to 3 + (12)$ transfer reaction in which the vertex 
$1+ 2 \to (12)$ is described by the form factor $(|\phi_{3}>)^{*}$. }
\label{fig_diagram(23)3Coulsingle1}
\end{figure}

\begin{figure}[tbp] 
\includegraphics[width=4.0in,height=4.5in]{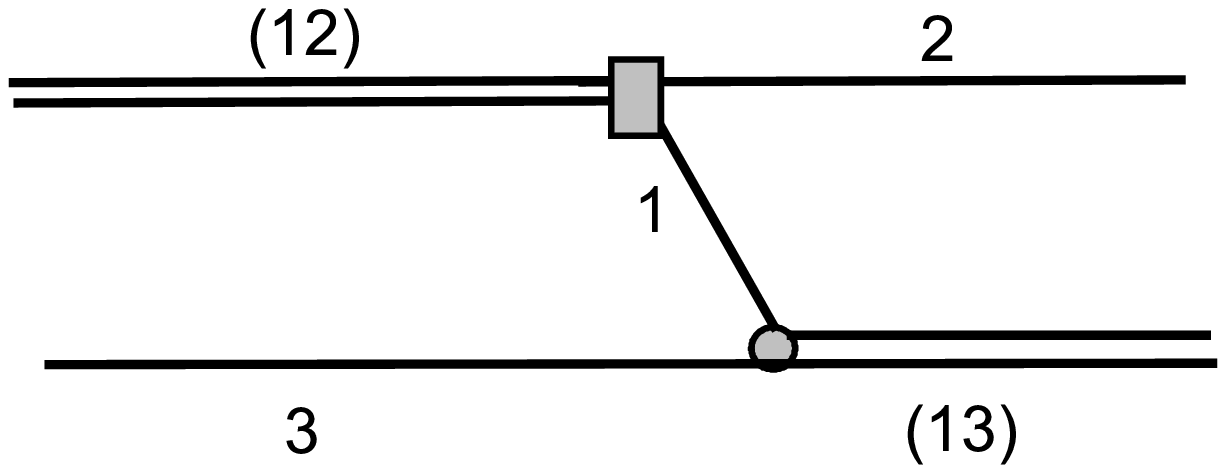}
  \caption{Diagram describing the $(12) + 3 \to 2 + (13)$ transfer reaction in which the vertex 
$(12) \to 1+2$ is described by the form factor $|\phi_{3}>$. }
\label{fig_diagram(12)2Coulsingle1}
\end{figure}

Thus the effective potentials for $\alpha=3$ or $\beta=3$, 
$\alpha \not= \beta$, which are given by the sum of two terms, see diagrams in Figs \ref{fig_pole(23)3} and \ref{fig_(12)2Coul}, can be presented by the single diagrams using 
Eq. (\ref{phinu31}). These diagrams are shown in Figs \ref{fig_diagram(23)3Coulsingle1} and \ref{fig_diagram(12)2Coulsingle1}, in which the vertex $(12) \leftrightarrow  1+2$ corresponds to the Coulomb-modified vertex form factor $|\phi_{3}>$.

\begin{figure}[tbp] 
\includegraphics[width=4.5in,height=4.5in]{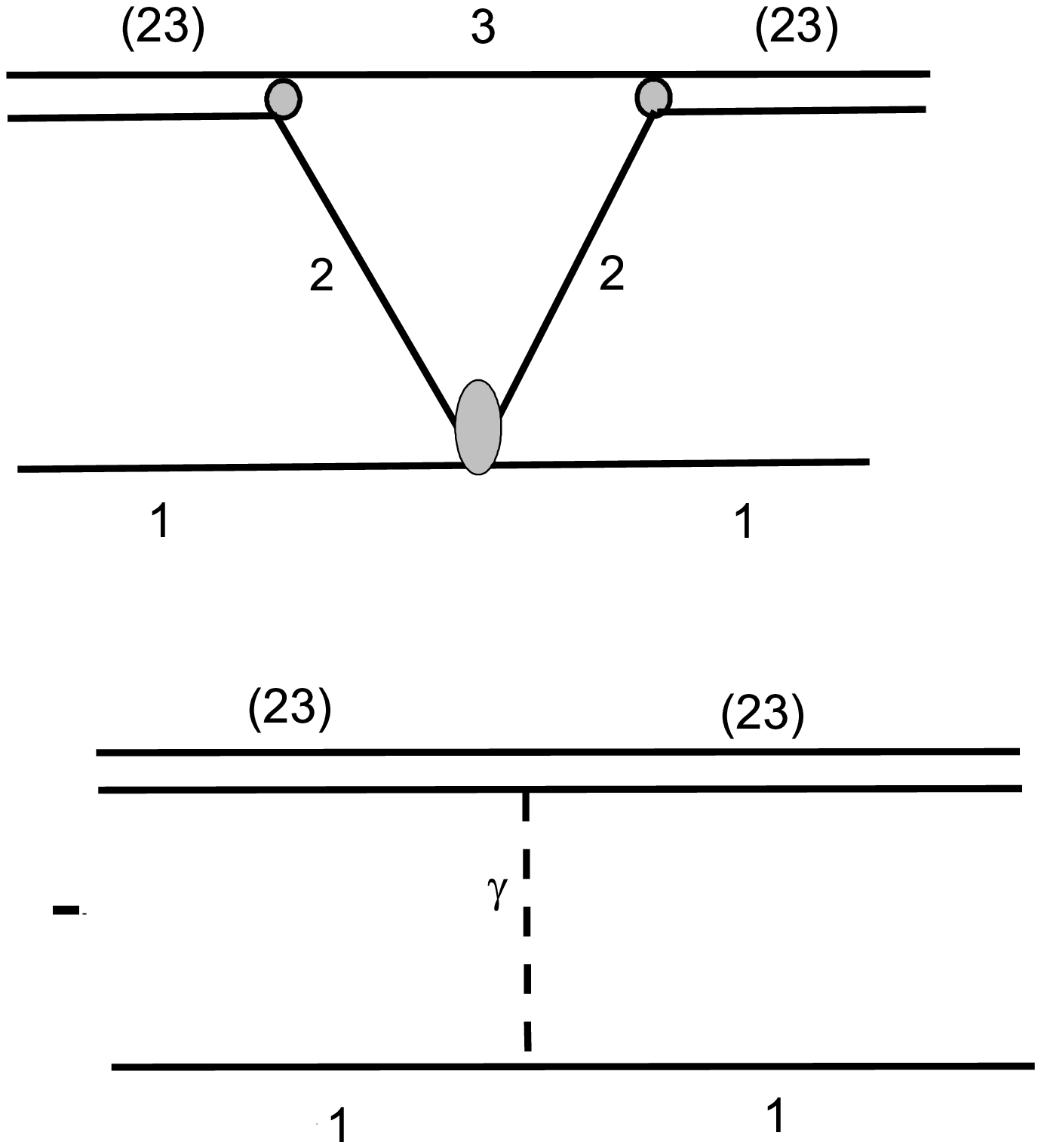}
  \caption{Triangle diagram describing the $(23) + 1 \to (23) + 1$ elastic scattering via the $1+2$ Coulomb scattering. Dashed bulb is the $1+2$ is the Coulomb elastic scattering amplitude. Also is shown the subtracted diagram 
corresponding to the photon exchange between particle $1$ and the c.m. of the bound state $(23)$.}
\label{fig_trianglediagram(23)1}
\end{figure}

\begin{figure}[tbp] 
\includegraphics[width=4.0in,height=4.5in]{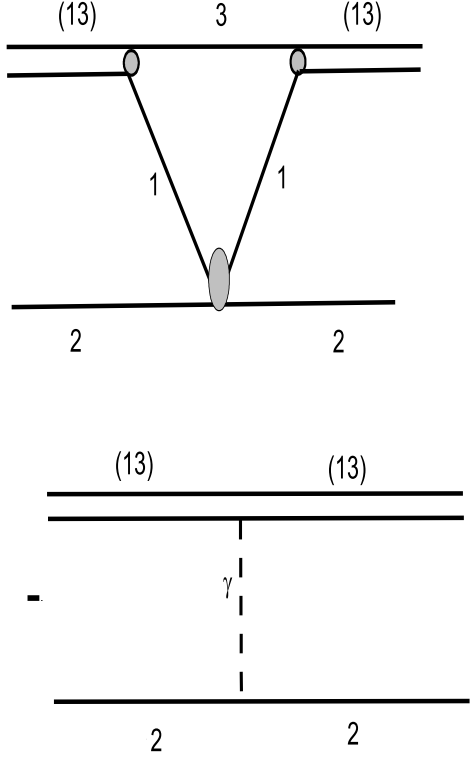}
  \caption{Triangle diagram describing the $(13) + 2 \to (13) + 2$ elastic scattering via the $1+2$ Coulomb scattering. Dashed bulb is the $1+2$  Coulomb elastic scattering amplitude.  Also is shown the subtracted diagram corresponding to the photon exchange between particle $2$ and the c.m. of the bound state $(13)$}
\label{fig_trianglediagram(13)2}
\end{figure}

\begin{figure}[tbp] 
\includegraphics[width=4.5in,height=4.5in]{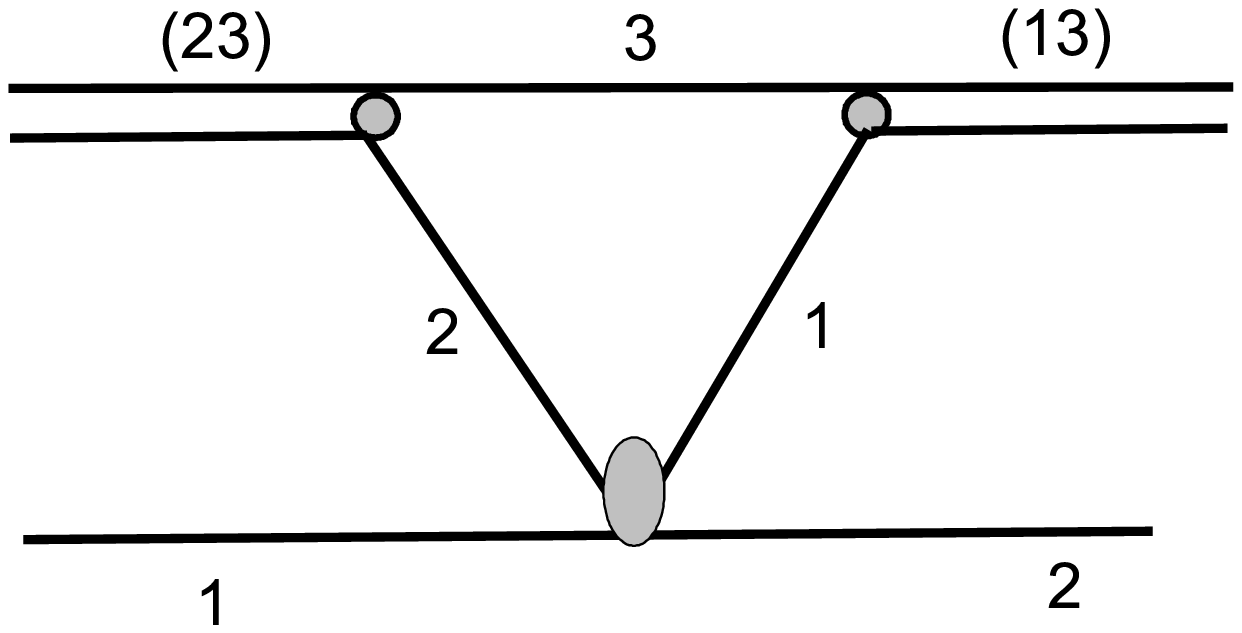}
  \caption{Exchange triangle diagram describing the $(23) + 1 \to (13) + 2$ transfer reaction. Dashed bulb is the $1+2$ is the Coulomb elastic scattering amplitude.}
\label{fig_exchangetriangle(23)2}
\end{figure}

The second bracket in Eq. (\ref{Vbaonsh11}) is the triangular diagram without the Born term describing the elastic scattering  with the four-ray vertex corresponding to the Coulomb scattering of particles $1$ and $2$, see Figs \ref{fig_trianglediagram(23)1} and \ref{fig_trianglediagram(13)2} . Finally, the last bracket in Eq. (\ref{Vbaonsh11}) is the triangular exchange diagram in Fig. \ref{fig_exchangetriangle(23)2} with the four-ray vertex corresponding to the Coulomb scattering of particles $1$ and $2$. Also we can add the diagrams corresponding to the inverse reactions.
  
Let us return now to Eq. (\ref{AGCCoulm1}). The main advantage of the AGS formalism with pure separable potentials is that it allows one to reduce the three-particle Faddeev equations to effective two-particle ones. The three-body Green functions in this approach are absorbed into the effective potentials.

\subsection{AGS equation with separable potentials for $A(d,p)B$ reaction amplitude}
\label{AGSseppotent1}

Now I rewrite Eqs. (\ref{AGCCoulm1}) for the $A(d,p)B$ reaction amplitude, which is coupled to the reaction amplitudes in other channels, assuming that the left-hand-side amplitude is ONES:
\begin{widetext}
\begin{align}
&{\cal T}_{p\,A}^{NC(R)}({\rm {\bf q}}_{p}^{C(R)(-)},{\rm {\bf q}}_{A}^{C(R)(+)};\,E_{+}) = {\cal V}_{p\,A}^{NC(R)}({\rm {\bf q}}_{p}^{C(R)(-)},{\rm {\bf q}}_{A}^{C(R)(+)};E_{+})                   \nonumber\\
&+ \sum\limits_{i=A,\,p,\,n}\int\,\frac{{\rm d}{\rm {\bf p}}_{i}}{(2\,{\pi})^{3}}\,{\cal {\tilde V}}_{p\,i}^{NC(R)}({\rm {\bf q}}_{p}^{C(R)(-)},{\rm {\bf p}}_{i}^{C(R)(-)};E_{+})\,\frac{1}{E_{+} +\varepsilon_{i} - p_{i}^{2}/(2\,{\cal M}_{i})}\,{\cal T}_{i\,A}^{NC(R)}({\rm {\bf p}}_{i}^{C(R)(-)},{\rm {\bf q}}_{A}^{C(R)(+)};\,E_{+}).
\label{AGCCoulpd1}
\end{align} 
\end{widetext}
Note that the initial channel $d(p\,n) + A$ is denoted by the free particle $A$, while the final channel $p+B(n\,A)$ by $p$. Also in the c.m. ${\rm {\bf p}}_{d}= - {\rm {\bf p}}_{A}$ and ${\rm {\bf p}}_{p}= - {\rm {\bf p}}_{B},$ $\,\varepsilon_{p} \equiv \varepsilon_{nA}$, $\varepsilon_{n} \equiv \varepsilon_{pA}$ and
$\varepsilon_{A} \equiv \varepsilon_{pn}$,  $\,{\cal M}_{p} \equiv {\cal M}_{pB}= m_{p}\,m_{nA}/{\cal M}$, $\,{\cal M}_{n} \equiv {\cal M}_{nF}=m_{n}\,m_{pA}/{\cal M}$ and  ${\cal M}_{A} \equiv {\cal M}_{dA}=m_{A}\,m_{pn}/{\cal M}$, $\,{\cal M} = m_{p} + m_{n} + m_{A}$. 
Also $g_{p} \equiv g_{nA},\,g_{A} \equiv g_{pn}$ and $g_{n} \equiv g_{pF}$ are the form factors,

The ONES effective potential ${\cal V}_{p\,A}^{NC(R)}({\rm {\bf q}}_{p}^{C(R)(-)},{\rm {\bf q}}_{A}^{C(R)(+)};E_{+})$  in Eq. (\ref{AGCCoulpd1}) (the first term on the right-hand-side)  is
\begin{align}
&{\cal V}_{p\,A}^{NC(R)}({\rm {\bf q}}_{p}^{C(R)(-)},{\rm {\bf q}}_{A}^{C(R)(+)};E_{+})   
= <\psi_{{\rm {\bf q}}_{p}}^{C(R)(-)}\,g_{p}\big|G_{0}(E_{+})     \nonumber\\
& + G_{0}(E_{+})\,T_{3}^{C(R)}(E_{+})\,G_{0}(E_{+})\big|g_{A}\,\psi_{{\rm {\bf q}}_{A}}^{C(R)(+)}>.
\label{effpotVpA1}
\end{align}
The other three effective potentials under the integral sign are:
\begin{widetext}
\begin{align}
{\cal {\tilde V}}_{p\,A}^{NC(R)}({\rm {\bf q}}_{p}^{C(R)(-)},{\rm {\bf p}}_{A}^{C(R)(-)};E_{+}) = <\psi_{{\rm {\bf q}}_{p}}^{C(R)(-)}\,g_{p}\big|\big[G_{0}(E_{+})     
 + G_{0}(E_{+})\,T_{3}^{C(R)}(E_{+})\,G_{0}(E_{+})\big]\,S_{A}(E_{+}- {\overline K}_{A})\big|g_{A}\,\psi_{{\rm {\bf p}}_{A}}^{C(R)(-)}>,
\label{effpotVpA2}
\end{align}
\end{widetext}
\begin{align}
&{\cal {\tilde V}}_{p\,p}^{NC(R)}({\rm {\bf q}}_{p}^{C(R)(-)},{\rm {\bf p}}_{p}^{C(R)(-)};E_{+}) \nonumber\\
&= <\psi_{{\rm {\bf q}}_{p}}^{C(R)(-)}\,g_{p}\big|G_{0}(E_{+})\big(T_{n}^{C(R)}(E_{+})S_{p}(E_{+}- {\overline K}_{p}) \nonumber\\
&- U_{n}^{C(R)}\big)G_{0}(E_{+})\big|g_{p}\,\psi_{{\rm {\bf p}}_{p}}^{C(R)(-)}>,
\label{effpotVpp2}
\end{align}
\begin{align}
&{\cal {\tilde V}}_{p\,n}^{NC(R)}({\rm {\bf q}}_{p}^{C(R)(-)},{\rm {\bf p}}_{n};E_{+}) \nonumber\\
&= <\psi_{{\rm {\bf q}}_{p}}^{C(R)(-)}\,g_{p}\big|\big[G_{0}(E_{+}) + G_{0}(E_{+})T_{3}^{C(R)}G_{0}(E_{+})\big] \nonumber\\
&\times S_{n}(E_{+}- {\overline K}_{n})\big|g_{n}\,\psi_{{\rm {\bf p}}_{n}}>,
\label{effpotVpn12}
\end{align}
${\overline K}_{A}$ is the kinetic energy operator of the relative motion of $A$ and $d$,
$S_{A}(E_{+}- {\overline K}_{A})\,|{\rm {\bf p}}_{A}>= |{\rm {\bf p}}_{A}>S_{A}(E_{+}- \frac{p_{A}^{2}}{2\,{\cal M}_{A}})$.

 The effective potential ${\cal V}_{p\,A}^{NC(R)}({\rm {\bf q}}_{p}^{C(R)(-)},{\rm {\bf q}}_{A}^{C(R)(+)};E_{+})$ is given by the sum of the diagrams in Figs. \ref{fig_pole(23)2} and \ref{fig_exchangetriangle(23)2} sandwiched by the Coulomb distorted waves. The ONES effective potential 
${\cal V}_{p\,p}^{NC(R)}({\rm {\bf q}}_{p}^{C(R)(-)},{\rm {\bf q}}_{p}^{C(R)(-)};E_{+})$ is given by the difference of the diagrams in Fig. \ref{fig_trianglediagram(13)2} sandwiched by the Coulomb distorted waves. 
The screened Coulomb-Born $d-A$ potential $U_{dA}^{C(R)}$ is subtracted from the triangular amplitude to compensate for the most singular term coming from the Born term of the $p-A$ scattering $T$-matrix.                
Finally, the ONES effective potential ${\cal V}_{p\,n}^{NC(R)}({\rm {\bf q}}_{p}^{C(R)(-)},{\rm {\bf q}}_{n};E)$ is described by the diagram  in Fig. \ref{fig_diagram(12)2Coulsingle1} or by the sum of the diagram in Fig. \ref{fig_(12)2Coul} sandwiched by the Coulomb distorted waves. $\,T_{3}^{C(R)} \equiv T_{pA}^{C(R)}$  is the off-shell Coulomb $\,p-A$ $\,T$-matrix generated by the screened $\,p-A$ Coulomb potential. I remind that in these diagrams $1=A,\,2=p$ and $3=n$.

\subsection{AGS equations for the sub-Coulomb $(d,p)$ reactions on heavier nuclei}
\label{AGSsepsubCoul1}

Let us consider now the application of the AGS equations for the sub-Coulomb $(d,p)$ reactions on heavier targets for which the Coulomb penetrability factors are very small. Because all the amplitudes and effective potentials are sandwiched by the Coulomb scattering wave functions containing the penetrability factors each effective potential and amplitude in Eq. (\ref{AGCCoulpd1}) also becomes very small. Hence, one can replace in Eq. (\ref{AGCCoulpd1}) the reaction amplitude
${\cal T}_{p\,A}^{NC(R)}({{\rm {\bf p}}_{p}^{C(R)(-)},{\rm {\bf q}}_{A}^{C(R)(+)};E_{+}})$ under the integral
sign with $i=p$ on the right-hand-side 
by the effective potential ${\cal V}_{p\,A}^{NC(R)}({{\rm {\bf p}}_{p}^{C(R)(-)},{\rm {\bf q}}_{A}^{C(R)(+)};E})$. At sub-Coulomb $(d,p)$ reactions on heavier target the elastic $d-A$ scattering is dominated by the Coulomb one. To simplify Eq. (\ref{AGCCoulpd1}) the elastic scattering amplitude ${\cal T}_{A\,A}^{NC(R)}({{\rm {\bf p}}_{A}^{C(R)(-)},{\rm {\bf q}}_{A}^{C(R)(+)};E_{+}})$ in the term with $i=A$ is replaced by the HOES pure Coulomb $d-A$ elastic scattering amplitude ${\cal {\tilde T}}_{A\,A}^{C(R)}({{\rm {\bf p}}_{A}^{C(R)(-)},{\rm {\bf q}}_{A}^{C(R)(+)};E_{+}}) $ generated by the channel Coulomb potential $U_{dA}^{C(R)}$ from which the Born Coulomb term is subtracted.

The amplitude ${\tilde{\cal T}}_{A\,A}^{(C(R))}({{\rm {\bf p}}_{A}^{C(R)(-)},{\rm {\bf q}}_{A}^{C(R)(+)};E_{+}})$ is given by the integral term in Eq. (B.10) \cite{muk2012}. Its operator takes the form
\begin{align}
{\tilde {\cal T}}_{A\,A}^{C(R)}(z) \equiv {\tilde {\cal T}}_{d\,d}^{C(R)}(z)= U_{dA}^{C(R)}\,{\hat G}_{dA}^{C(R)}(z)\,U_{dA}^{C(R)}.
\label{tildeT1}
\end{align}
Here, ${\hat G}_{dA}^{C(R)}(z) = (z + \varepsilon_{pn} - {\overline K}_{dA}- U_{dA}^{C(R)})^{-1}$ is the two-body $d-A$ Coulomb Green function, ${\overline K}_{dA}$ is the kinetic energy operator of the $d-A$ relative motion.
  
Then Eq. (\ref{AGCCoulpd1}) reduces to 
\begin{widetext}
\begin{align}
&{\cal T}_{p\,A}^{NC(R)}({\rm {\bf q}}_{p}^{C(R)(-)},{\rm {\bf q}}_{A}^{C(R)(+)};\,E_{+}) = {\cal V}_{p\,A}^{NC(R)}({\rm {\bf q}}_{p}^{C(R)(-)},{\rm {\bf q}}_{A}^{C(R)(+)};E_{+})                    \nonumber\\
& + \int\,\frac{{\rm d}{\rm {\bf p}}_{A}}{(2\,{\pi})^{3}}{\cal {\tilde V}}_{p\,A}^{NC(R)}({\rm {\bf q}}_{p}^{C(R)(-)},{\rm {\bf p}}_{A}^{C(R)(-)};E_{+})\,\frac{1}{E_{+} +\varepsilon_{A} - p_{A}^{2}/(2\,{\cal M}_{A}) +i0}\,{\tilde {\cal T}}_{A\,A}^{C(R)}({{\rm {\bf p}}_{A}^{C(R)(-)},{\rm {\bf q}}_{A}^{C(R)(+)};E_{+}})
\nonumber\\
&+ \int\,\frac{{\rm d}{\rm {\bf p}}_{p}}{(2\,{\pi})^{3}}{\cal {\tilde V}}_{p\,p}^{NC(R)}({\rm {\bf q}}_{p}^{C(R)(-)},{\rm {\bf p}}_{p}^{C(R)(-)};E_{+})\,\frac{1}{E_{+} +\varepsilon_{p} - p_{p}^{2}/(2\,{\cal M}_{p}) +i0}\,{\cal V}_{p\,A}^{NC(R)}({{\rm {\bf p}}_{p}^{C(R)(-)},{\rm {\bf q}}_{A}^{C(R)(+)};E_{+}})
\nonumber\\
& +\int\,\frac{{\rm d}{\rm {\bf p}}_{n}}{(2\,{\pi})^{3}}{\cal {\tilde V}}_{p\,n}^{NC(R)}({\rm {\bf q}}_{p}^{C(R)(-)},{\rm {\bf p}}_{n};E_{+})\,\frac{1}{E_{+} +\varepsilon_{n} - p_{n}^{2}/(2\,{\cal M}_{n}) +i0}\,{\tilde {\cal T}}_{n\,A}^{NC(R)}({{\rm {\bf p}}_{n},{\rm {\bf q}}_{A}^{C(R)(+)};E_{+}}).
\label{AGCCoulpd2}
\end{align} 
\end{widetext}

Thus for the sub-Coulomb $A(d,p)B$ reactions on the heavier nuclei the AGS coupled equations are reduced to one expression (\ref{AGCCoulpd2}) in which the only unknown amplitude is the ONES 
amplitude ${\cal T}_{p\,A}^{NC(R)}({\rm {\bf q}}_{p}^{C(R)(-)},{\rm {\bf q}}_{A}^{C(R)(+)};\,E_{+})$
on the left-hand-side.
My goal is to analyze the peripheral character of the expression (\ref{AGCCoulpd2}) for the sub-Coulomb $(d,p)$ reactions rather than solving it.
 
At sub-Coulomb energies, due to the presence of the Coulomb scattering wave functions in the $d-A$ and $p-B$ channels, the $(d,p)$ reactions are peripheral and are contributed by a few smallest partial waves. Peripheral character in the momentum space means that in the intermediate states the integration momenta $p_{i}$ do not deviate much from the on-shell values $q_{i}$. 
For the peripheral $(d,p)$ reactions the dominant contribution comes from $r_{nA} > 1/\kappa_{nA}$; ${\rm {\bf r}}_{ij}$ is the radius-vector connecting particles $i$ and $j$, and $\kappa_{ij}$ is the bound-state wave number of the bound state $(ij)$. In the momentum space it is equivalent to the dominant contribution of the momenta $p_{ij} < \kappa_{ij}$, where ${\rm {\bf p}}_{ij}$ is the momentum conjugated to ${\rm {\bf r}}_{ij}$. 

In the DWBA for the peripheral $A(d,p)B$ reaction the $B=(nA)$ bound-state wave function can be replaced by its asymptotic tail whose amplitude is the asymptotic normalization coefficient (ANC) $C_{nA}$.
Then the DWBA cross section is proportional to the $C_{nA}^{2}$. Normalizing the DWBA cross section to the experimental one we can determine the ANC what constitutes the ANC method \cite{muk90,reviewpaper}. The question is whether the amplitude of the sub-Coulomb $A(d,p)B$ reaction calculated using the AGS equation (\ref{AGCCoulpd2}) is peripheral and can be parametrized in terms of the ANC $C_{nA}$.

Let us begin with the effective potential ${\cal V}_{p\,A}^{NC(R)}({\rm {\bf q}}_{p}^{C(-)},{\rm {\bf q}}_{A}^{C(+)};E_{+})$, which is the first term on the right-hand-side of Eq. (\ref{AGCCoulpd2}). 
Now I show that it can be expressed in terms of the sub-Coulomb DWBA amplitude plus next order term. The proof requires a few transformations.  
\begin{align}
&{\cal V}_{p\,A}^{NC(R)}({\rm {\bf q}}_{p}^{C(R)(-)},{\rm {\bf q}}_{A}^{C(R)(+)};E_{+})   
= <\psi_{{\rm {\bf q}}_{p}}^{C(R)(-)}\,g_{p}\big|G_{0}(E_{+})     \nonumber\\
& + G_{0}(E_{+})\,T_{3}^{C(R)}(E_{+})\,G_{0}(E_{+})\big|g_{A}\,\psi_{{\rm {\bf q}}_{A}}^{C(R)(+)}> \nonumber\\
&= <\psi_{{\rm {\bf q}}_{p}}^{C(R)(-)}\,g_{p}\big|G_{pA}^{C(R)}(E_{+})\big|g_{A}\,\psi_{{\rm {\bf q}}_{A}}^{C(R)(+)}>.
\label{VpAtr1}
\end{align}
For the Coulomb Green function $G_{pA}^{C(R)}(z)$ of the particles $p$ and $A$ interacting via the screened Coulomb potential $V_{n}^{C(R)} \equiv V_{pA}^{C(R)}$ in the three-body space  I use the post-transformation:
\begin{align}
&G_{pA}^{C(R)}(z) = \frac{1}{z- K - V_{pA}^{C(R)}} = \frac{1}{z- K - \Delta V_{pA}^{C(R)(post)} - U_{p}^{C(R)}}                        \nonumber\\
&= G_{pB}^{C(R)}(z)\,\Big( 1 + \Delta V_{pA}^{C(R)(post)}\,G_{pA}^{C(R)}(z)\Big). 
\label{GnpCR1}
\end{align}
Here, $G_{pB}^{C(R)}(z)=(z- K- U_{p}^{C(R)})^{-1}$, 
$\,\,U_{p}^{C(R)} \equiv U_{pB}^{C(R)}$,  $\,\,\Delta V_{pA}^{C(R)(post)}= V_{pA}^{C(R)} - U_{pB}^{C(R)}$.
Then 
\begin{widetext}
\begin{align}
&{\cal V}_{p\,A}^{NC(R)}({\rm {\bf q}}_{p}^{C(R)(-)},{\rm {\bf q}}_{A}^{C(R)(+)};E_{+}) 
= <\psi_{{\rm {\bf q}}_{p}}^{C(R)(-)}\,g_{p}\big|G_{pA}^{C(R)}(E_{+})\big|g_{A}\,\psi_{{\rm {\bf q}}_{A}}^{C(R)(+)}>                                                           \nonumber\\          
&= <\psi_{{\rm {\bf q}}_{p}}^{C(R)(-)}\,g_{p}\big|G_{pB}^{C(R)}(E_{+})\,\Big( 1 + \Delta V_{pA}^{C(R)(post)}\,G_{pA}^{C(R)}(E_{+})\Big)\big|g_{A}\,\psi_{{\rm {\bf q}}_{A}}^{C(R)(+)}>  \nonumber\\
&= <\psi_{{\rm {\bf q}}_{p}}^{C(R)(-)}\,\varphi_{p}\big| 1 + \Delta V_{pA}^{C(R)(post)}\,G_{pA}^{C(R)}(E_{+})\big|g_{A}\,\psi_{{\rm {\bf q}}_{A}}^{C(R)(+)}>.
\label{VpADWBA1}
\end{align}
\end{widetext}
Here the equation
\begin{align}
&<\psi_{{\rm {\bf q}}_{p}}^{C(R)(-)}\,g_{p}\big|G_{pB}^{C(R)}(E_{+})     \nonumber\\
&= - <\psi_{{\rm {\bf q}}_{p}}^{C(R)(-)}\,g_{p}\big|\frac{1}{\varepsilon_{p} + K_{\alpha}} 
= <\psi_{{\rm {\bf q}}_{p}}^{C(R)(-)}\,\varphi_{p}\big|.
\label{intwf1}
\end{align}
is used. Also $\varphi_{p} \equiv \varphi_{nA}$ is the $B=(nA)$ bound-state wave function. 
Now, instead of the post transformation, I use in Eq. (\ref{VpADWBA1}) the prior transformation of  $G_{pA}^{C(R)}(z)$:
\begin{align}
&G_{pA}^{C(R)}(z) =  \frac{1}{z- K - \Delta V_{pA}^{C(R)(prior)} - U_{dA}^{C(R)}}                        \nonumber\\
&=\Big( 1 + G_{pA}^{C(R)}(z)\,\Delta V_{pA}^{C(R)(prior)}\Big)\,G_{dA}^{C(R)}(z), 
\label{GnACR1}
\end{align}
where $\,\,G_{dA}^{C(R)}(z)=(z- K- U_{dA}^{C(R)})^{-1}$, $\,\,\Delta V_{pA}^{C(R)(prior)}=
V_{pA}^{C(R)} - U_{dA}^{C(R)}$.
Then
\begin{widetext}
\begin{align}
&{\cal V}_{p\,A}^{NC(R)}({\rm {\bf q}}_{p}^{C(R)(-)},{\rm {\bf q}}_{A}^{C(R)(+)};E_{+})
= <\psi_{{\rm {\bf q}}_{p}}^{C(R)(-)}\,g_{p}\big|G_{pA}^{C(R)}(E_{+})\big|g_{A}\,\psi_{{\rm {\bf q}}_{A}}^{C(R)(+)}>                    \nonumber\\                                      
&= <\psi_{{\rm {\bf q}}_{p}}^{C(R)(-)}\,\varphi_{p}\big| 1 + \Delta V_{pA}^{C(R)(post)}\Big(G_{pA}^{C(R)}(E_{+})\,\Delta V_{pA}^{C(R)(prior)} + 1\Big)\,G_{dA}^{C(R)}(E_{+}) \big|g_{A}\,\psi_{{\rm {\bf q}}_{A}}^{C(R)(+)}>                                                         \nonumber\\
&= <\psi_{{\rm {\bf q}}_{p}}^{C(R)(-)}\,\varphi_{p}\big|g_{A}\,\psi_{{\rm {\bf q}}_{A}}^{C(R)(+)}>
+ <\psi_{{\rm {\bf q}}_{p}}^{C(R)(-)}\,\varphi_{p}\big|\Delta V_{pA}^{C(R)(post)} + \Delta V_{pA}^{C(R)(post)}\,G_{pA}^{C(R)}(E_{+})\,\Delta V_{pA}^{C(R)(prior)}\,\big|\varphi_{A}\,\psi_{{\rm {\bf q}}_{A}}^{C(R)(+)}>
.
\label{VpADWBA2}
\end{align}
\end{widetext} 

Taking into account that 
\begin{align}
<\varphi_{p}\big|g_{A}>= <\varphi_{A} \big|V_{A}\big|\varphi_{A}>
\label{varphig1}
\end{align}
one can reduce Eq. (\ref{VpADWBA2}) to
\begin{widetext}
\begin{align}
&{\cal V}_{p\,A}^{NC(R)}({\rm {\bf q}}_{p}^{C(R)(-)},{\rm {\bf q}}_{A}^{C(R)(+)};E_{+}) 
=<\psi_{{\rm {\bf q}}_{p}}^{C(R)(-)}\,g_{p}\big|G_{pA}^{C(R)}(E_{+})\big|g_{A}\,\psi_{{\rm {\bf q}}_{A}}^{C(R)(+)}>                                                            \nonumber\\
&= {\cal T}_{pA}^{DW}({\rm {\bf q}}_{p}^{C(R)(-)},{\rm {\bf q}}_{A}^{C(R)(+)})
+ <\psi_{{\rm {\bf q}}_{p}}^{C(R)(-)}\,\varphi_{p}\big| \Big(V_{pA}^{C(R)} - U_{pB}^{C(R)}\Big)\,G_{pA}^{C(R)}(E_{+})\,\Big(V_{pA}^{C(R)} - U_{dA}^{C(R)}\Big)\,\big|\varphi_{pn}\,\psi_{{\rm {\bf q}}_{A}}^{C(R)(+)}>,
\label{VpADWBAp1}
\end{align}
where the post-form of the sub-Coulomb DWBA amplitude is
\end{widetext}
\begin{align}
&{\cal T}_{pA}^{DW}({\rm {\bf q}}_{p}^{C(R)(-)},{\rm {\bf q}}_{A}^{C(R)(+)};E_{+})     \nonumber\\
& = <\psi_{{\rm {\bf q}}_{p}}^{C(R)(-)}\,\varphi_{nA}\big|V_{pn} + V_{pA}^{C(R)} - U_{pB}^{C(R)}\big|  \varphi_{pn}\,\psi_{{\rm {\bf q}}_{A}}^{C(R)(-)}>.
\label{TDWBApost1}
\end{align}
If I would change the order of application of the prior and post transformations  of $G_{pA}^{C(R)}(E_{+})$, then the effective potential ${\cal V}_{p\,A}^{NC(R)}({\rm {\bf q}}_{p}^{C(R)(-)},{\rm {\bf q}}_{A}^{C(R)(+)};E_{+})$ can be expressed in terms of the prior-form DWBA amplitude. 

Now I can easily show that the amplitude ${\cal T}_{pA}^{DW}({\rm {\bf q}}_{p}^{C(R)(-)},{\rm {\bf q}}_{A}^{C(R)(+)};E_{+})$ 
is peripheral for the sub-Coulomb $A(d,p)B$ on heavier targets for which the Coulomb parameters in the initial and final states $\eta_{q_{A}},\,\eta_{q_{p}} >>1$, where $\eta_{q_{A}}= (Z_{d}\,Z_{A}/137)\mu_{dA}/q_{A}$, $\,\,q_{A} \equiv q_{dA}$, and $\eta_{q_{p}}= (Z_{p}\,Z_{B}/137)\mu_{pB}/q_{p}$, $\,\,q_{p} \equiv q_{pB}$, $Z_{p}= Z_{d}=1$ and $Z_{B}=Z_{A}$. First, I introduce the partial wave decomposition of the DWBA amplitude which can be schematically written as 
\begin{align}
{\cal T}_{pA}^{DW}(q_{p\,l_{p}}^{C(R)},\,q_{A\,l_{A}}^{C(R)};E_{+}) = <\psi_{q_{p}\,l{p}}^{C(R)}\big|O(r)\big|\psi_{q_{A}\,l_{A}}^{C(R)}>,
\label{quasclass1}
\end{align}
where   $\psi_{q_{A}\,l_{A}}^{C(R)}$ and $\psi_{q_{p}\,l{p}}^{C(R)}$ are the Coulomb scattering wave functions in the initial and final states; $l_{A} \equiv l_{dA}$ ($l_{p} \equiv l_{pB}$)  is the relative orbital angular momentum in the initial $d+A$ (final $p+B$) channel.
All other functions of the  matrix elements, except for the partial Coulomb distorted waves, are absorbed into $O(r)$.
Now it is convenient to use the quasiclassical approach for the Coulomb partial waves \cite{alder}:
\begin{align}
\psi_{q\,l}  \sim \sqrt{\Big[\frac{f(r)}{q^{2}}\Big]^{-1/4}}\,\sin\,\phi,
\label{psiquas1}
\end{align}  
\begin{align}
f(r) =q^{2} - \frac{l\,\eta}{r} - \frac{l(l+1)}{r^{2}},
\label{f1}
\end{align}
\begin{align}
\phi = \frac{\pi}{4} + \int\limits_{r_{0}}^{r}\,{\rm d}\,\sqrt{[f(r)]^{1/2}}.
\label{phi1}
\end{align}  
$r_{0}$ is the classical turning point determined by the condition: $f(r_{0}) =0$. $r_{0}$ increases with increasing of $\eta$.  Thus from the classical approach, which is valid at large Coulomb parameter $\eta$, follows that the dominant contribution to the Coulomb partial wave give $r >r_{0}$, while the internal distances $r< r_{0}$, which are located in the classically forbidden region, give negligible contribution. Hence, any matrix element sandwiched by the partial Coulomb distorted waves, is peripheral. For example, for the ${}^{208}{\rm Pb}(d,p){}^{209}{\rm Pb}$ reaction at  $E_{dA}=5$ MeV (the Coulomb barrier is $V_{CB}= 12.2$ MeV) and the head-on collision $l_{dA}=0$ in the initial channel $\,r_{0}= 23.6$ fm. Such a large $r_{0}$ makes the reaction amplitude both peripheral and small. Head-on collision is dominant because for $l_{dA} >0$ $r_{0}$ increases decreasing the reaction amplitude. The Rutherford trajectory at head on-collisions is peaked backward. Hence the proton differential cross section generated by the amplitude ${\cal T}_{pA}^{DW}({\rm {\bf q}}_{p}^{C(R)(-)},{\rm {\bf q}}_{A}^{C(R)(+)}; E_{+})$ is backward peaked. 

\begin{figure}[tbp] 
\includegraphics[width=4.0in,height=4.5in]{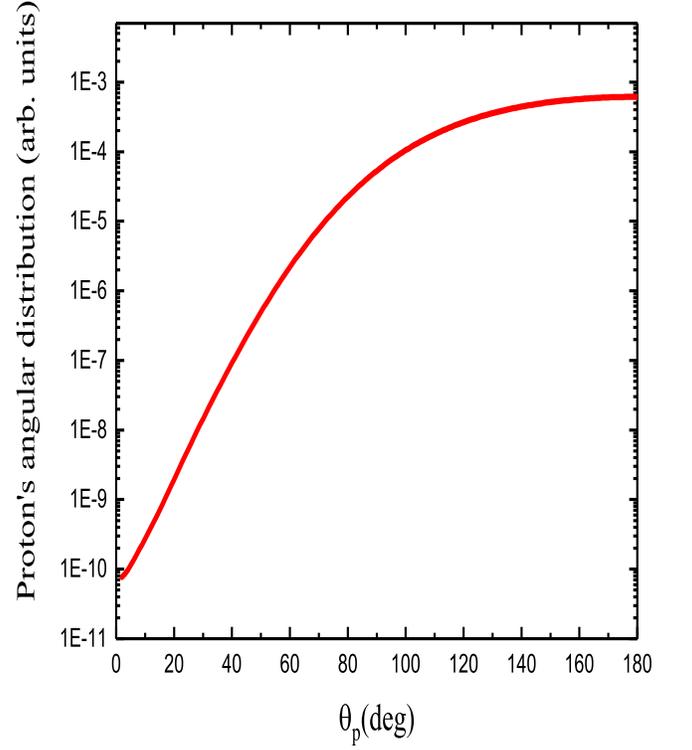}
  \caption{Proton's angular distribution for the ${}^{208}{\rm Pb}(d,p){}^{209}{\rm Pb}$ reaction at  $E_{dA}=5$ MeV populating the $2g_{9/2}$ ground state in ${}^{209}{\rm Pb}$. $\theta_{p}$ is the proton scattering angle in the c.m. of the reaction. }
\label{fig_angdistr208Pb}
\end{figure}

To demonstrate it in Fig. \ref{fig_angdistr208Pb} is shown the proton's angular distribution in the direct ${}^{208}{\rm Pb}(d,p){}^{209}{\rm Pb}$ reaction at $E_{dA}=5$ MeV calculated using the DWBA FRESCO code \cite{FRESCO}. It is a sub-Coulomb reaction because the Coulomb barrier is $\,V_{CB} \approx 12.2$ MeV and the Coulomb parameter in the initial state is $\eta_{dA} =8.16$. Thus this process demonstrates a perfect example of the sub-Coulomb reaction with large Coulomb parameters. The proton's angular distribution, as explained, has a pronounced backward peak. In the calculations the Reid soft-core potential for the deuteron bound state and standard Woods-Saxon for the neutron $2g_{9/2}$ bound state in $\,{}^{209}{\rm Pb}$ are used.
However, the details of the adopted potentials are not important because the backward peak is an universal pattern of the angular distribution of sub-Coulomb direct transfer reactions on nuclei with higher charges.  

In summarizing the analysis of the effective potential ${\cal V}_{pA}({\rm {\bf q}}_{p}^{C(R)(-)},{\rm {\bf q}}_{A}^{C(R)(+)}; E_{+})$, I remind the proved essential results: for the sub-Coulomb $(d,p)$ reactions the effective potential ${\cal V}_{pA}({\rm {\bf q}}_{p}^{C(R)(-)},{\rm {\bf q}}_{A}^{C(R)(+)}; E_{+})$ whose mechanism is described by the sum of the pole and triangular exchange diagrams in Figs. \ref{fig_pole(23)2} and \ref{fig_exchangetriangle(23)2}, correspondingly, is dominantly contributed by the DWBA amplitude ${\cal T}_{pA}^{DW}({\rm {\bf q}}_{p}^{C(R)(-)},{\rm {\bf q}}_{A}^{C(R)(+)};E_{+})$. The second term in Eq. (\ref{VpADWBAp1}) is significantly smaller then the DWBA amplitude because, after the spectral decomposition of the Green function, it contains four penetrability factors versus two in the DWBA amplitude.  
If the energies in the initial and final states are well below the Coulomb barrier then the  amplitude 
${\cal T}_{pA}^{DW}({\rm {\bf q}}_{p}^{C(R)(-)},{\rm {\bf q}}_{A}^{C(R)(+)}; E_{+})$ of the
$A(d,p)B$ is peripheral and parametrized in terms of $C_{nA}$, where $C_{nA}$ is the ANC of the $(nA)$ bound state. The differential cross section generated by ${\cal T}_{pA}^{DW}({\rm {\bf q}}_{p}^{C(R)(-)},{\rm {\bf q}}_{A}^{C(R)(+)}; E_{+})$ is backward peaked at sub-Coulomb energies on heavier targets \cite{Knutson}.

Let us return to Eq. (\ref{AGCCoulpd2}).
The integrand of the second term on the right-hand-side of this equation contains the effective potential
${\cal {\tilde V}}_{p\,A}^{NC(R)}({\rm {\bf q}}_{p}^{C(R)(-)},{\rm {\bf p}}_{A}^{C(R)(-)};E_{+})$, which is expressed
in terms of the HOES DWBA ${\cal T}_{pA}^{DW}({\rm {\bf q}}_{p}^{C(R)(-)},{\rm {\bf p}}_{A}^{C(R)(+)}; E_{+})$. The matrix element of the partial wave HOES DWBA amplitude  written in the quasiclassical approach is peripheral and contains the factor $e^{-|\zeta|\,\pi/2}$ \cite{alder}, where $\zeta= \eta_{q_{p}} - \eta_{p_{A}}$. 
Hence, at large Coulomb parameter $\eta_{q_{p}}$ the dominant contribution in the integral over $p_{A}$ comes from minimal $\zeta$. From the previous discussion it is evident that ${\cal {\tilde V}}_{p\,A}^{NC(R)}({\rm {\bf q}}_{p}^{C(R)(-)},{\rm {\bf p}}_{A}^{C(R)(-)};E_{+})$ is peripheral with regard to the bound-state wave function $\varphi_{p} \equiv \varphi_{nA}$ and is parametrized in terms of ANC $C_{nA}$. Hence, the second term of Eq. (\ref{AGCCoulpd2}) is also peripheral and is parametrized in terms of ANC $C_{nA}$. 
 
The same is true for the third term on the right-hand-side of Eq. (\ref{AGCCoulpd2}), which contains ${\cal{\tilde V}}_{p\,A}^{NC(R)}({\rm {\bf p}}_{p}^{C(-)},{\rm {\bf q}}_{A}^{C(+)};E_{+})$.  This amplitude again is dominantly contributed by the HOES DWBA amplitude 
${\cal T}_{p\,A}^{DW}({\rm {\bf p}}_{p}^{C(-)},{\rm {\bf q}}_{A}^{C(+)};E_{+})$. Evidently that the DWBA amplitude is peripheral and parametrized in terms of the ANC $C_{nA}$ what is also true for the whole  third term.
 
Now let us consider the fourth term.  This term contains only two Coulomb penetrability factors corresponding to the initial and final states because in the intermediate states we have $n+F$ channel in which the channel Coulomb interaction is absent. The fourth team describes the two-step reaction $d+ A \to n + F \to B + p$.
The integrand of the fourth term contains two amplitudes, ${\cal {\tilde V}}_{p\,n}^{NC(R)}({\rm {\bf q}}_{p}^{C(R)(-)},{\rm {\bf p}}_{n};E_{+})$ and ${\cal T}_{n\,A}^{NC(R)}({{\rm {\bf p}}_{n},{\rm {\bf q}}_{A}^{C(R)(+)};E_{+}})$ describing the first and second steps, correspondingly.
The effective potential ${\cal {\tilde V}}_{p\,n}^{NC(R)}({\rm {\bf q}}_{p}^{C(-)},{\rm {\bf p}}_{n};E_{+})$ is the amplitude of the proton transfer reaction $A(d,n)F$, where $F=(p\,A)$, and is described  by the diagram in Fig. \ref{fig_diagram(23)3Coulsingle1}.
The amplitude $\,\,{\cal T}_{n\,A}^{NC(R)}({{\rm {\bf p}}_{n},{\rm {\bf q}}_{A}^{C(R)(+)};E_{+}})$ of the $\,F(n,\,B)p$ reaction contains the Coulomb distorted wave in the initial channel $\,d+A$. This Coulomb distorted wave makes the reaction amplitude at the sub-Coulomb energy $\,E_{dA}$ peripheral and small and can be approximated by its effective potential $\,\,{\cal V}_{n\,A}^{NC(R)}({{\rm {\bf p}}_{n},{\rm {\bf q}}_{A}^{C(R)(+)};E})$ described by the pole diagram shown in Fig. \ref{fig_diagram(12)2Coulsingle1}. The notations of the particles are the same as in the previous cases.

\begin{figure}[tbp] 
\includegraphics[width=4.0in,height=4.5in]{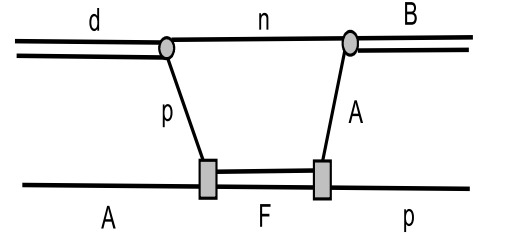}
  \caption{Rectangular diagram describing the two-step process $d + A \to B + p$. }
\label{fig_rectangdiagram}
\end{figure}

Then the rectangular diagram describing the fourth term (without the Coulomb distorted waves in the initial and final states of the reaction, which do not affect the location of the singularities of the diagram) is shown in Fig. \ref{fig_rectangdiagram}.

\begin{figure}[tbp] 
\includegraphics[width=4.0in,height=4.5in]{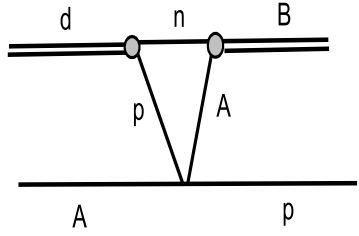}
  \caption{Triangular diagram describing the reaction $d + A \to B + p$ obtained from the rectangular diagram in Fig. \ref{fig_rectangdiagram} by contracting the line $F$. }
\label{fig_trianglediagramdABp}
\end{figure}

To find its nearest to the physical region singularity in the $cos({\rm {\bf q}}_{p} \cdot {\rm {\bf q}}_{d})$ plane (${\rm {\bf q}}_{d} = - {\rm {\bf q}}_{A}$), which governs the angular distribution of the cross section generated by this diagram, one can contract the line $F$ in the rectangular diagram in Fig. \ref{fig_rectangdiagram} reducing it to the triangular diagram in Fig. \ref{fig_trianglediagramdABp} , which is the skeleton diagram of the rectangular diagram. The nearest to the physical region singularity of the ONES triangular diagram, and, hence, of the rectangular diagram, generated by the propagators (all the vertices are taken to be constant) is located in the $cos({\rm {\bf q}}_{p} \cdot {\rm {\bf q}}_{d})$ plane at 
\begin{align}
z_{t} = - \frac{m_{d}\,m_{B}}{2\,m_{n}^{2}}\,\frac{ (\frac{m_{n}}{m_{d}})^{2}q_{d}^{2} + (\frac{m_{n}}{m_{B}})^{2}q_{p}^{2} + (\kappa_{pn} + \kappa_{nA})^{2} }{q_{d}\,q_{p} }  < -1.
\label{z0t1}
\end{align}

This singularity is located quite far away from the border of the physical region $cos({\rm {\bf q}}_{p} \cdot {\rm {\bf q}}_{d})=-1$. The nearest to the physical region singularity of the ONES amplitude of the pole diagram in Fig. \ref{fig_pole(23)2} (the notations of the particles are the same as in the previous cases) is
\begin{align}
z_{p} = \frac{m_{d}}{2\,m_{p}}\,\frac{q_{p}^{2} + (\frac{m_{p}}{m_{d}})^{2}q_{d}^{2} + \kappa_{pn}^{2}}{q_{d}\,q_{p}} >1.
\label{z0t1}
\end{align}
It is located on the opposite site of the unphysical region but much closer to the border of the physical region $cos({\rm {\bf q}}_{p} \cdot {\rm {\bf q}}_{d})=1$ than the singularity of the triangular diagram. 

\begin{figure}[tbp] 
\includegraphics[width=4.0in,height=4.5in]{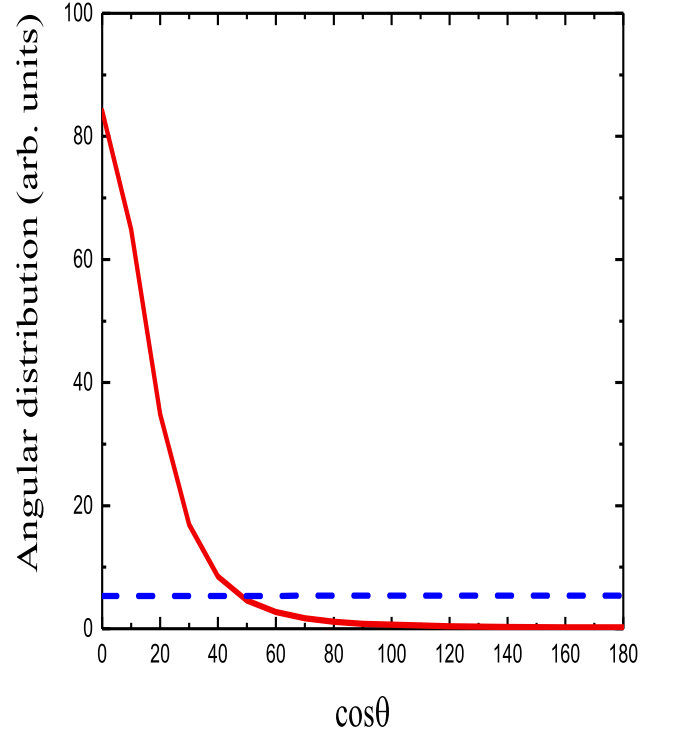}
  \caption{Angular distributions for the  $d + A \to B + p$ reaction at $E_{dA}= 5$ MeV generated by $({\rm cos}{\rm{\theta}} - z_{t})^{-2}$, dashed blue line, and by $({\rm cos}{\rm {\theta}} - z_{p})^{-2}$, solid red line. }
\label{fig_angdistrzpzt1}
\end{figure}

As an example, I consider the sub-Coulomb  ${}^{208}{\rm Pb}(d,\,p){}^{209}{\rm Pb}$ reaction at  $E_{dA}= 5$ MeV. For this case we get $z_{t}=-432.048$ and $z_{p}=1.11$. 
These singularities govern the angular distributions generated by the corresponding diagrams. 
In Fig. \ref{fig_angdistrzpzt1} are shown the angular distributions generated by $({\rm cos}{\rm{\theta}} - z_{t})^{-2}$ and $({\rm cos}{\rm {\theta}} - z_{p})^{-2}$. As we see, the angular distribution generated by the pole singularity has pronounced forward peak while the triangular singularity produces absolutely flat angular distribution. The folding of the amplitude of the pole diagram with the Coulomb distorted waves in the initial and final states converts the forward peak into the backward one  because of the dominant head-on collision while the angular distribution generated by the rectangular diagram sandwiched with the Coulomb distorted waves remains flat.  

Therefore, one can neglect the contribution of the fourth term in Eq. (\ref{AGCCoulpd2}) at the backward proton angles compared to the first three terms on the right-hand-side of Eq. (\ref{AGCCoulpd2}). 

Because the second and third terms contain four penetrability factors each, they are smaller than the first term, ${\cal V}_{p\,A}^{NC(R)}({\rm {\bf q}}_{p}^{C(-)},{\rm {\bf q}}_{A}^{C(+)};E_{+})$.
Thus it is shown that for the sub-Coulomb $A(d,p)B$ reactions on heavier nuclei the AGS amplitude ${\cal T}_{p\,A}^{NC(R)}({\rm {\bf q}}_{p}^{C(R)(-)},{\rm {\bf q}}_{A}^{C(R)(+)};\,E_{+})$ with separable potentials is well approximated by the post form of the sub-Coulomb DWBA amplitude:
\begin{align}
&{\cal T}_{p\,A}^{NC(R)}({\rm {\bf q}}_{p}^{C(R)(-)},{\rm {\bf q}}_{A}^{C(R)(+)};\,E_{+}) \nonumber\\
&\approx {\cal V}_{p\,A}^{NC(R)}({\rm {\bf q}}_{p}^{C(R)(-)},{\rm {\bf q}}_{A}^{C(R)(+)};E_{+}) \nonumber\\
&= {\cal T}_{p\,A}^{DW}({\rm {\bf q}}_{p}^{C(R)(-)},{\rm {\bf q}}_{A}^{C(R)(+)};E_{+}).
\label{subCoulDWBA1}
\end{align}

Since the sub-Coulomb DWBA amplitude is peripheral and parametrized in terms of the ANC $C_{nA}$ of the bound state $(nA)$, the same is also the case for the AGS  $A(d,p)B$ reaction amplitude 
${\cal T}_{p\,A}^{NC(R)}({\rm {\bf q}}_{p}^{C(R)(-)},{\rm {\bf q}}_{A}^{C(R)(+)};\,E_{+})$.
For better accuracy one can add to the DWBA amplitude the second and third terms of the right-hand-side of Eq. (\ref{AGCCoulpd2}), which can become important when energy $E_{dA}$ increases but still below the Coulomb barrier.

\section{AGS equations with general local potentials}
\label{AGSgeneralpotential1}

In this section the AGS equations are written for general forms of the two-body local potentials rather than for nonlocal separable potentials.  
I briefly describe the derivation of these equations because it will be used in the next section where the AGS equations are modified by including the optical potentials. The AGS equations can be derived directly from the equations for the transition operator (\ref{tildeUCRplus1}):
\begin{widetext}
\begin{align}
&U_{\beta\,\alpha}^{NC(R)(+)}(z) = {\Delta{\overline V}_{\beta}^{(R)}} +
{\Delta{\overline V}_{\beta}^{(R)}}\,G^{(R)}(z)\,{\Delta{\overline V}_{\alpha}^{(R)}}             \nonumber\\
&= {\Delta{\overline V}_{\beta}^{(R)}} + \Big[{V}_{\alpha}^{N}\,G^{(R)}(z) + {V}_{\gamma}^{N}\,G^{(R)}(z)   + \Delta {\overline V}_{\beta}^{C(R)}\,G^{(R)}(z)\,\Big]\,\Delta {\overline V}_{\alpha}^{(R)}       \nonumber\\
& = \Delta{\overline V}_{\beta}^{(R)} + \Big[ {V}_{\alpha}^{N}\,{\overline G}_{\alpha}^{(R)}(z)\,\big(\Delta{\overline V}_{\alpha}^{(R)}\,G^{(R)}(z) + 1 \big) +
{V}_{\gamma}^{N}\,{\overline G}_{\gamma}^{(R)}(z)\,\big(\Delta{\overline V}_{\gamma}^{(R)}\,G^{(R)}(z)\, + 1 \big)          \nonumber\\
& + {\Delta {\overline V}_{\beta}^{C(R)}}\,{\overline G}_{\beta}^{(R)}(z)\big({\Delta{\overline V}_{\beta}^{(R)}}\,G^{(R)}(z) +1 \big) \Big]{\Delta{\overline V}_{\alpha}^{(R)}} 
\nonumber\\ 
&={\Delta{\overline V}_{\beta}^{(R)}} + \sum\limits_{\nu}\,\Big[{\overline {\delta}_{\beta\,\nu}}\,{V}_{\nu}^{N} + \delta_{\beta\,\nu}\,{\Delta {{\overline V}}^{C(R)}_{\nu}} \Big]\,{\overline G}_{\nu}^{(R)}(z)\,U_{\nu\,\alpha}^{NC(R)(+)}(z),
\label{Uplus11}
\end{align}
\end{widetext}
\begin{align}
{\overline G}_{\alpha}^{(R)}(z) = \frac{1}{z - K - V_{\alpha} - U_{\alpha}^{C(R)}}.
\label{GalphaR1}
\end{align}
  
To derive the coupled equations for the transition operator  the potential  ${\Delta{\overline V}}_{\alpha}^{(R)}$ has been split 
into three terms: two nuclear potentials $V_{\beta}^{N}$ and $V_{\gamma}^{N}$, and one Coulomb term
${\Delta {{\overline V}}^{C(R)}_{\alpha}}$. This allows us to express $U_{\beta\,\alpha}^{NC(R)(+)}(z)$ in terms of the three components, $U_{\nu\,\alpha}^{NC(R)(+)}(z)$, $\,\nu= \beta,\,\gamma,\,\alpha$.

The ONES reaction amplitude is given  by the matrix element from $U_{\beta\,\alpha}^{NC(R)(+)}(z)$ taken between the initial and final physical states:
\begin{widetext}
\begin{align}
&{\cal T}_{\beta\,\alpha}^{NC(R)}({\rm {\bf q}}_{\beta}^{C(R)(-)},\,{\rm {\bf q}}_{\alpha}^{C(R)(+)};E_{+}) 
= <\psi_{ {\rm {\bf q}}_{\beta}}^{C(R)(-)}\,\varphi_{\beta}^{(R} \Big|U_{\beta\,\alpha}^{NC(R)(+)}(E_{+})
 \Big| \varphi_{\alpha}^{(R)}\,\psi_{ {\rm {\bf q}}_{\alpha}}^{C(R)(+)}>,
\label{Mba1}
\end{align}
\end{widetext}
where $\psi_{{\rm {\bf q}}_{\alpha}}^{C(R)(+)}$ and $\psi_{{\rm {\bf q}}_{\beta}}^{C(R)(-)}$ are the Coulomb scattering wave functions in the initial and final states  calculated for the screened channel Coulomb potentials $U_{\alpha}^{C(R)}$ and $U_{\beta}^{C(R)}$, correspondingly. 

Instead of the transition operator $U_{\beta\,\alpha}^{NC(R)(+)}(z)$ one may consider the transition operator
\begin{align}
U_{\beta\,\alpha}^{NC(R)}(z) = {\overline \delta}_{\beta\,\alpha}\,[{\overline G}_{\alpha}^{(R)}(z)]^{-1} +
U_{\beta\,\alpha}^{NC(R)(+)}(z).
\label{tildeUba1}
\end{align}
Then $\,U_{\beta\,\alpha}^{NC(R)}(z)$ satisfies the equation
\begin{align}
&U_{\beta\,\alpha}^{NC(R)}(z) = {\overline \delta}_{\beta\,\alpha}\,\Big([{\overline G}_{\alpha}^{(R)}(z)]^{-1} + V_{\alpha}^{N} \Big) +  \delta_{\beta\,\alpha}\,{\Delta {\overline V}_{\beta}^{C(R)}}  \nonumber\\
& +  \sum\limits_{\nu}\Big[{\overline \delta}_{\beta\,\nu}\,{V}_{\nu}^{N} + \delta_{\beta\,\nu}\,{\Delta {{\overline V}}^{C(R)}_{\nu}} \Big]
\,{\overline G}_{\nu}^{(R)}(z)\,U_{\nu\,\alpha}^{NC(R)}(z) .
\label{Uba2}
\end{align}

Equations (\ref{Uba2}) were derived in \cite{deltuva2005}. 
Note that the ONES matrix elements from $\,U_{\beta\,\alpha}^{NC(R)(+)}(z)\,$ and $U_{\beta\,\alpha}^{NC(R)}(z)$, in which the final state is physical, coincide:
\begin{align} 
&<\psi_{ {\rm {\bf q}}_{\beta}}^{C(R)(-)}\,\varphi_{\beta}^{(R)} \Big|U_{\beta\,\alpha}^{NC(R)(+)}(E_{+}) \Big| \varphi_{\alpha}^{(R)}\,\psi_{ {\rm {\bf p}}_{\alpha}}^{C(R)(+)}>   \nonumber\\
&= <\psi_{ {\rm {\bf q}}_{\beta}}^{C(R)(-)}\,\varphi_{\beta}^{(R)} \Big|U_{\beta\,\alpha}^{NC(R)}(E_{+}) \Big| \varphi_{\alpha}^{(R)}\,\psi_{ {\rm {\bf p}}_{\alpha}}^{C(R)(+)}>
\label{Uba12}
\end{align}
because
\begin{align}
 [{\overline G}_{\alpha}^{(R)}(E_{+})]^{-1}\,\big|\varphi_{\alpha}^{(R)}\,\psi_{ {\rm {\bf q}}_{\alpha}}^{C(R)(+)}>=0.
\label{GbetaR1physstate1} 
\end{align}

After having derived the Faddeev equations for the transition operators $U_{\beta\,\alpha}^{NC(R)(+)}(z)$ we can write down the Faddeev 
equations in the AGS form for the reaction amplitude. For the separable potentials the Faddeev three-body equations are reduced to the two-body AGS equations. For general potentials it is not the case. When writing the AGS equations for the general potentials  Eq. (\ref{Uplus11}) is used in which one needs to introduce the spectral decomposition of the Green functions ${\overline G}_{\nu}^{(R)}(z)$.
This spectral decomposition contains both two-body and three-body terms. Here, when deriving the AGS equations for the reaction amplitudes, I neglect the three-body terms in the spectral decomposition of ${\overline G}_{\nu}^{(R)}(z)$, that is, the contribution from the three-body continuum in the intermediate states is neglected. 
Thus I use the spectral decomposition of the Green functions ${\overline G}_{\nu}^{(R)}(z)$: 
\begin{align}
{\overline G}_{\nu}^{(R)}(z) =  \int \frac{ {\rm d} {\rm {\bf p}}_{\nu} }{(2\,\pi)^{3}} \frac{|\,{\rm {\bf p}}_{\nu}^{C(R)(-)}\,\varphi_{\nu}^{(R)}><\varphi_{\nu}^{(R)}\,{\rm {\bf p}}_{\nu}^{C(R)(-)}|}{z + \varepsilon_{\nu} -  \frac{p_{\nu}^{2}}{2\,{\cal M}_{\nu}} }.
\label{Galphaspectrdecomp1}
\end{align}
Neglecting the contribution from the three-body continuum in the spectral decomposition of the channel Green functions ${\overline G}_{\nu}^{(R)}(z)$  allows us to derive the two-particle Faddeev equations in the AGS form in which the effective potentials are expressed in terms of the DWBA amplitudes for the sub-Coulomb transfer reactions.
Also only one bound state is taken into account in each channel. The extension for a few bound states is straightforward and will be demonstrated in section  \ref{generalizedAGSCDCC1}.

Taking the matrix elements from the left- and right-hand-sides of Eq. (\ref{Uplus11}) and using the spectral decomposition (\ref{Galphaspectrdecomp1})  we get 
\begin{widetext}
\begin{align}
& {\cal T}_{\beta\,\alpha}^{NC(R)}({\rm {\bf q}}_{\beta}^{C(R)(-)},\,{\rm {\bf q}}_{\alpha}^{C(R)(+)}; E_{+})   
 = {\cal T}^{DW}_{\beta\,\alpha}({\rm {\bf q}}_{\beta}^{C(R)(-)},\,{\rm {\bf q}}_{\alpha}^{C(R)(+)}; E_{+})  \nonumber\\
&  + \sum\limits_{\nu}\, \int{\frac{ {\rm d} {\rm {\bf p}}_{\nu}  }{ (2\,\pi)^{3} }   } \frac{{\cal {\tilde T}}_{\beta\,\nu}^{DW}({\rm {\bf q}}_{\beta}^{C(R)(-)},\,{\rm {\bf p}}_{\nu}^{C(R)(-)}; E_{+}) \,{\cal T}_{\nu\,\alpha}^{NC(R)}({\rm {\bf p}}_{\nu}^{C(R)(-)},\,{\rm {\bf q}}_{\alpha}^{C(R)(+)}; E_{+}) }{E_{+} +\varepsilon_{\nu} - \frac{p_{\nu}^{2}}{2\,{\cal T}_{\nu}} }. 
\label{Teq1}
\end{align}
\end{widetext}
These are desired Faddeev equations written as two-particle AGS ones. It is worth mentioning that in these equations the explicit coupling of the transfer reactions and elastic scattering amplitudes are taken into account while the contribution from the breakup channel is neglected.  In contrast, in the well-known CDCC method \cite{Austern} or more simplified adiabatic distorted waves (ADWA) \cite{Johnson}  the coupling of the specific transfer reaction channel and the breakup channel is taken into account but the explicit coupling to other transfer reaction channels and elastic scattering is neglected. 
Thus while the AGS two-particle equations with separable potentials are obtained without any approximation,
when one uses general local potentials it is not the case. 
 
The reaction amplitudes and effective potentials in Eq. (\ref{Teq1}) are
\begin{align} 
&{\cal T}_{\beta\,\alpha}^{NC(R)}({\rm {\bf q}}_{\beta}^{C(R)(-)},\,{\rm {\bf q}}_{\alpha}^{C(R)(+)}; E_{+})                                                                             \nonumber\\
&= <\psi_{ {\rm {\bf q}}_{\beta}}^{C(R)(-)}\,\varphi_{\beta}^{(R)} \Big|U_{\beta\,\alpha}^{NC(R)(+)}(E_{+}) \Big| \varphi_{\alpha}^{(R)}\,\psi_{ {\rm {\bf q}}_{\alpha}}^{C(R)(+)}>,  
\label{Ubamatrelem1}
\end{align}
\begin{align}
&{\cal T}^{DW}_{\beta\,\alpha}({\rm {\bf q}}_{\beta}^{C(R)(-)},\,{\rm {\bf q}}_{\alpha}^{C(R)(+)}; E_{+})            \nonumber\\
& =  <\psi_{ {\rm {\bf q}}_{\beta}}^{C(R)(-)}\,\varphi_{\beta}^{(R)} \Big|{\Delta{\overline V}_{\beta}^{(R)}} \Big| \varphi_{\alpha}^{(R)}\,\psi_{ {\rm {\bf q}}_{\alpha}}^{C(R)(+)}>,
\label{TDW1}
\end{align}
\begin{align}
&{\cal {\tilde T}}_{\beta\,\alpha}^{DW}({\rm {\bf q}}_{\beta}^{C(R)(-)},\,{\rm {\bf p}}_{\alpha}^{C(R)(-)}; E_{+})     \nonumber\\
& = <{\rm {\bf q}}_{\beta}^{C(R)(-)}\,\varphi_{\beta}^{(R)} \big| V_{\alpha}^{N} \big| \varphi_{\alpha}^{(R)}\,{\rm {\bf p}}_{\alpha}^{C(R)(-)}>,
\label{TDW2}
\end{align}
\begin{align}
&{\cal {\tilde T}}_{\beta\,\beta}^{DW}({\rm {\bf q}}_{\beta}^{C(R)(-)},\,{\rm {\bf p}}_{\beta}^{C(R)(-)}; E_{+})    \nonumber\\
&=  <{\rm {\bf q}}_{\beta}^{C(R)(-)}\,\varphi_{\beta}^{(R)}\big |{\Delta {\overline V}_{\beta}^{C(R)}}\big |\varphi_{\beta}^{(R)}\,{\rm {\bf p}}_{\beta}^{C(R)(+)}>,
\label{TDW3}
\end{align}
\begin{align}
&{\cal {\tilde T}}_{\beta\,\gamma}^{DW}({\rm {\bf q}}_{\beta}^{C(R)(-)},\,{\rm {\bf p}}_{\gamma}^{C(R)(-)}; E_{+})      \nonumber\\
&= <{\rm {\bf q}}_{\beta}^{C(R)(-)}\,\varphi_{\beta}^{(R)}\big| V_{\gamma}^{N} \big|\varphi_{\gamma}^{(R)}\,{\rm {\bf p}}_{\gamma}^{C(R)(-)}>.
\end{align}
Here ${\cal T}^{DW}_{\beta\,\alpha}({\rm {\bf q}}_{\beta}^{C(R)(-)},\,{\rm {\bf q}}_{\alpha}^{C(R)(+)}; E_{+})$  is ONES sub-Coulomb DWBA reaction amplitude in the post form. ${\cal {\tilde T}}_{\beta\,\alpha}^{DW}({\rm {\bf q}}_{\beta}^{C(R)(-)},\,{\rm {\bf p}}_{\alpha}^{C(R)(-)}; E_{+})$ is HOES DWBA sub-Coulomb reaction amplitude with the transition operator $V_{\alpha}^{N}$. 
${\cal {\tilde T}}_{\beta\,\beta}^{DW}({\rm {\bf  q}}_{\beta}^{C(R)(-)},\,{\rm {\bf p}}_{\beta}^{C(R)(-)}; E_{+})$ is the HOES DWBA elastic scattering amplitude with pure Coulombic transition operator ${\Delta {\overline V}_{\beta}^{C(R)}}$. $\,{\cal {\tilde T}}_{\beta\,\gamma}^{DW}({\rm {\bf q}}_{\beta}^{C(R)(-)},\,{\rm {\bf p}}_{\gamma}^{C(R)(-)}; E_{+})$ is the HOES DWBA sub-Coulomb reaction amplitude with the transition operator $V_{\gamma}^{N}$.

\subsection{Sub-Coulomb $(d,p)$ reactions}
\label{SubCoulomb1}

Equations (\ref{Teq1}) are very convenient for the analysis of the peripheral character of the sub-Coulomb $A(d,p)B$ reactions because they contain the Coulomb distorted waves in the initial and final states of the matrix elements, which have crucial importance for the sub-Coulomb reactions.  
The transition operator in this case satisfies equation
\begin{widetext}
\begin{align}
&U_{p\,A}^{NC(R)(+)}(z)                             
={\Delta{\overline V}_{p}^{(R)}} + \sum\limits_{i=A,p,n}\,\Big[{\overline {\delta}_{p\,i}}\,{V}_{i}^{N} + \delta_{p\,i}\,{\Delta {{\overline V}}^{C(R)}_{i}} \Big]\,{\overline G}_{i}^{(R)}(z)\,U_{i\,A}^{NC(R)(+)}(z).
\label{Uplpd1}
\end{align}
\end{widetext}
The channel indexes $p,\,A,\,n$ correspond to the channels $p+ B(nA),\,d(pn)+A,\,n + F(pA)$, correspondingly,  while the potential $V_{p}  \equiv V_{nA}$, $\,V_{A} \equiv V_{pn}$ and $V_{n}^{(R)} \equiv V_{pA}^{(R)}$,     
${\Delta {\overline V}}_{p}^{(R)} \equiv {\overline V}_{p}^{(R)} - U_{p}^{C(R)}$, $\,\,{\overline V}_{p}^{(R)}= V_{pn} + V_{pA}^{(R)}$, $\,\,U_{p}^{C(R)} \equiv U_{pB}^{C(R)}$, $\,\,{\Delta {{\overline V}}^{C(R)}_{p}} \equiv V_{pA}^{C(R)} - U_{pB}^{C(R)}$.  

Then the two-particle AGS equation for the $A(d,p)B$ reaction amplitude take the form
\begin{widetext}
\begin{align}
& {\cal T}_{p\,A}^{NC(R)}({\rm {\bf q}}_{p}^{C(R)(-)},\,{\rm {\bf q}}_{A}^{C(R)(+)}; E_{+})   
 = {\cal T}^{DW}_{p\,A}({\rm {\bf q}}_{p}^{C(R)(-)},\,{\rm {\bf q}}_{A}^{C(R)(+)}; E_{+})  \nonumber\\
&  + \sum\limits_{i=A,p,n}\,\int{\frac{ {\rm d} {\rm {\bf p}}_{i}  }{ (2\,\pi)^{3} }   } \frac{{\cal {\tilde T}}_{p\,i}^{DW}({\rm {\bf q}}_{p}^{C(R)(-)},\,{\rm {\bf p}}_{i}^{C(R)(-)}; E_{+}) \,{\cal T}_{i\,A}^{NC(R)}({\rm {\bf p}}_{i}^{C(R)(-)},\,{\rm {\bf q}}_{A}^{C(R)(+)}; E_{+}) }{E_{+} +\varepsilon_{i} - \frac{p_{i}^{2}}{2\,{\cal M}_{i}}}.
\label{TBa3}
\end{align}
\end{widetext}
Thus one of the important goals of the paper is achieved: the Faddeev equations in the two-particle AGS form with local potentials for the sub-Coulomb $A(d,p)B$ reactions has been derived. 
My goal is to demonstrate this equation is peripheral. 
Note that on the left-hand-side of Eq. (\ref{TBa3}) we have 
the ONES reaction amplitude ${\cal T}_{p\,A}^{NC(R)}({\rm {\bf q}}_{p}^{C(R)(-)},\,{\rm {\bf q}}_{A}^{C(R)(+)}; E_{+})$,
while under the integral sign the same reaction amplitude is HOES because the  momentum ${\rm {\bf p}}_{p}$
is the integration variable. 

The first term on the right-hand-side of Eq. (\ref{TBa3}) is the effective potential  
\begin{align}
&{\cal T}^{DW}_{p\,A}({\rm {\bf q}}_{p}^{C(R)(-)},\,{\rm {\bf q}}_{A}^{C(R)(+)}; E_{+})  
= <\psi_{ {\rm {\bf q}}_{p}}^{C(R)(-)}\,\varphi_{nA} \Big| V_{pn}  \nonumber\\
& + V_{pA}^{N} + V_{pA}^{C(R)} - U_{pB}^{C(R)} \Big| \varphi_{pn}\,\psi_{ {\rm {\bf q}}_{A}}^{C(R)(+)}>, 
\label{MDWpost1}
\end{align}
which is the ONES sub-Coulomb post-form of the DWBA   $A(d,p)B$ reaction amplitude. The effective potential in the second term on-the right-hand-side ($i=A$)
\begin{align}
&{\cal {\tilde T}}_{p\,A}^{DW}({\rm {\bf q}}_{p}^{C(R)(-)},\,{\rm {\bf p}}_{A}^{C(R)(-)}; E_{+})  \nonumber\\
&= <\psi_{{\rm {\bf q}}_{p}}^{C(R)(-)}\,\varphi_{nA} \big| V_{pn} \big| \varphi_{pn}\,\psi_{{\rm {\bf p}}_{A}}^{C(R)(-)}>,
\label{TtildepADW1}
\end{align}
is the HOES post-form of the DWBA $A(d,p)B$ reaction amplitude with $V_{pn}$ as the transition operator. 
The HOES elastic scattering amplitude ${\cal T}_{A\,A}^{NC(R)}({\rm {\bf p}}_{A}^{C(R)(-)},\,{\rm {\bf q}}_{A}^{C(R)(+)}; E_{+})$ under the integral in the second term 
at the sub-Coulomb energies can be replaced by the HOES pure Coulomb $d-A$ elastic scattering amplitude ${\cal {\tilde T}}_{A\,A}^{C(R)}({{\rm {\bf p}}_{A}^{C(R)(-)},{\rm {\bf q}}_{A}^{C(R)(+)};E_{+}}) $ generated by the channel Coulomb potential $U_{dA}^{C(R)}$ from which the Born Coulomb term is subtracted.
 
The effective potential in the third
term ($\,i=p\,$) on the right-hand-side
\begin{align}
&{\cal {\tilde T}}^{DW}_{p\,p}({\rm {\bf q}}_{p}^{C(R)(-)},\,{\rm {\bf p}}_{p}^{C(R)(-)}; E_{+})     \nonumber\\ 
&= <\psi_{ {\rm {\bf q}}_{p}}^{C(R)(-)}\,\varphi_{nA} \Big|V_{pA}^{C(R)} - U_{pB}^{C(R)}\Big| \varphi_{nA}\,\psi_{ {\rm {\bf p}}_{p}}^{C(R)(-)}> 
\label{MDWpppost1}
\end{align}
is the $p + B \to p +B$ HOES DWBA elastic scattering amplitude with the pure Coulombic transition operator $V_{pA}^{C(R)} - U_{pB}^{C(R)}$.
The reaction amplitude ${\cal T}_{p\,A}^{NC(R)}({\rm {\bf p}}_{p}^{C(R)(-)},\,{\rm {\bf q}}_{A}^{C(R)(+)}; E_{+})$ in the third term  at sub-Coulomb energies is small and in the leading order at  can be replaced by the HOES DWBA amplitude ${\cal T}_{p\,A}^{DW}({\rm {\bf p}}_{p}^{C(R)(-)},\,{\rm {\bf q}}_{A}^{C(R)(+)}; E_{+})$. 

Finally, the effective potential in the fourth term ($\,i=n\,$) is the HOES DWBA amplitude of the $n+F \to B + p$ reaction:
\begin{align}
&{\cal {\tilde  T}}^{DW}_{p\,n}({\rm {\bf q}}_{p}^{C(R)(-)},\,{\rm {\bf p}}_{n}; E_{+})     \nonumber\\
&= <\psi_{ {\rm {\bf q}}_{p}}^{C(R)(-)}\,\varphi_{nA} \Big| V_{pA}^{N} \Big| \varphi_{pA}^{C(R)}\,\psi_{ {\rm {\bf p}}_{n}}^{(0)}>. 
\label{MDWpn1}
\end{align} 
The reaction amplitude ${\cal T}_{n\,A}^{NC(R)}({\rm {\bf p}}_{n},\,{\rm {\bf q}}_{A}^{C(R)(+)}; E_{+})$ at the sub-Coulomb energies in the leading order can be replaced by the HOES DWBA reaction amplitude ${\cal T}_{n\,A}^{DW}({\rm {\bf p}}_{n},\,{\rm {\bf q}}_{A}^{C(R)(+)}; E_{+})$.

Then for the sub-Coulomb $A(d,p)B$ reaction the AGS Eq. (\ref{TBa3}) reduces to the expression for the AGS reaction amplitude:
\begin{widetext}
\begin{align}
& {\cal T}_{p\,A}^{NC(R)}({\rm {\bf q}}_{p}^{C(R)(-)},\,{\rm {\bf q}}_{A}^{C(R)(+)}; E_{+})   
 = {\cal T}^{DW}_{p\,A}({\rm {\bf q}}_{p}^{C(R)(-)},\,{\rm {\bf q}}_{A}^{C(R)(+)}; E_{+})  \nonumber\\
&  + \int{\frac{ {\rm d} {\rm {\bf p}}_{A}  }{ (2\,\pi)^{3} }   } \frac{{\cal {\tilde T}}_{p\,A}^{DW}({\rm {\bf q}}_{p}^{C(R)(-)},\,{\rm {\bf p}}_{A}^{C(R)(-)}; E_{+}) \,{\cal T}_{A\,A}^{NC(R)}({\rm {\bf p}}_{A}^{C(R)(-)},\,{\rm {\bf q}}_{A}^{C(R)(+)}; E_{+}) }{E_{+} +\varepsilon_{A} - \frac{p_{A}^{2}}{2\,{\cal M}_{A}}}                        \nonumber\\
& + \int{\frac{ {\rm d} {\rm {\bf p}}_{p}  }{ (2\,\pi)^{3} }   } \frac{{\cal {\tilde T}}_{p\,p}^{DW}({\rm {\bf q}}_{p}^{C(R)(-)},\,{\rm {\bf p}}_{p}^{C(R)(-)}; E_{+}) \,{\cal T}_{p\,A}^{DW}({\rm {\bf p}}_{p}^{C(R)(-)},\,{\rm {\bf q}}_{A}^{C(R)(+)}; E_{+}) }{E_{+} +\varepsilon_{p} - \frac{p_{p}^{2}}{2\,{\cal M}_{p}}}                                            \nonumber\\
&+ \int{\frac{ {\rm d} {\rm {\bf p}}_{n}}{ (2\,\pi)^{3}}}\,\frac{{\cal {\tilde T}}_{p\,n}^{DW}({\rm {\bf q}}_{p}^{C(R)(-)},\,{\rm {\bf p}}_{n}^{C(R)(-)}; E_{+}) \,{\cal T}_{n\,A}^{DW}({\rm {\bf p}}_{n},\,{\rm {\bf q}}_{A}^{C(R)(+)}; E_{+}) }{E_{+} +\varepsilon_{n} - \frac{p_{n}^{2}}{2\,{\cal M}_{n}}}
.
\label{subCoulTBa3}
\end{align}
\end{widetext}

Now, word by word I can repeat the end of subsection \ref{AGSsepsubCoul1}. The proof that for the sub-Coulomb $(d,p)$ reaction the AGS amplitude determined by expression (\ref{subCoulTBa3}) is peripheral is the same as in section{AGSsepsubCoul1} for the AGS equation with separable potentials.
Thus the sub-Coulomb $A(d,p)B$ reaction amplitude on heavier nuclei ${\cal T}_{p\,A}^{NC(R)}({\rm {\bf q}}_{p}^{C(R)(-)},{\rm {\bf q}}_{A}^{C(R)(+)};\,E_{+})$ with local potentials is peripheral and its normalization is determined by the ANC $C_{nA}$ of the bound state $(nA)$.\\

\section{AGS equations with included optical potentials}
\label{AGsopticalpotentials1}

Now I proceed to the section in which the Faddeev equations in the AGS form are generalized by
including the optical potentials. 
Usually the Faddeev equations were derived for real $V_{\alpha},\,\alpha=1,2,3$ potentials. For the first time the optical potentials in the AGS formalism were introduced in \cite{Sandhas1969} and in practice were used in \cite{alt2007} in the calculations of the ${}^{12}{\rm C}(d,p){}^{13}{\rm C}$ reactions using the AGS equations with separable potentials. The optical potential appeared because the excitation of the target ${}^{12}{\rm C}$ was taken into account. 
In \cite{deltuva2007,deltuva2009} the $V_{pB}$ optical potential was used when solving the AGS equations for the $A(d,p)B$ reactions. 

In this paper I present generalization of the Faddeev equations in the AGS form by including the optical potentials in addition to the basic real nuclear potentials $V_{\alpha}^{N}, \,\,\alpha=1,\,2,\,3$, which describe the interaction between the constituent particles $1,\,2$ and $3$. 
The optical potentials introduced in a way which is similar to the procedure used in the DWBA. The inclusion of the optical potentials in the transition operators will generate the optical model distorted waves in the initial and final channels of the reaction. These distorted waves are the solutions of the Schr\"odinger equation with the optical potentials, which are given by the sum of the nuclear optical and Coulomb channel potentials. 
Until now I introduced only the channel Coulomb potentials with the Coulomb distorted waves.  
Introducing the optical potentials allows one to express the effective potentials in the AGS equations in terms of the DWBA amplitudes. The goal is to derive the Faddeev equations in the 
two-particle AGS form with optical potentials. 

I start from the modified equation for the transition operator
\begin{align}
&{\tilde U}_{\beta\,\alpha}^{ONC(R)(+)}(z)            \nonumber\\
&= {\Delta{\overline V}_{\beta}^{ONC(R)}} +
{\Delta{\overline V}_{\beta}^{ONC(R)}}\,G^{(R)}(z)\,{\Delta{\overline V}_{\alpha}^{ONC(R)}}.
\label{UOR1}
\end{align}
Here 
\begin{align}
&{\Delta{\overline V}_{\alpha}^{ONC(R)}}= \Delta {\overline V}_{\alpha}^{ON} + \Delta {\overline V}_{\alpha}^{C(R)},
\label{VORalpha1}
\end{align}
\begin{align}
\Delta {\overline V}_{\alpha}^{ON} = V_{\beta}^{N} + V_{\gamma}^{N} - U_{\alpha}^{ON},
\label{ValphaON1}
\end{align}
where $U_{\alpha}^{ON}$ is the $\alpha$-channel nuclear optical potential describing the interaction between particle $\alpha$ and the c.m. of the bound state $(\beta\,\gamma)$.
$\Delta {\overline V}_{\alpha}^{C(R)}$ is given by Eq. (\ref{Ua11}). Superscript $ON$ means the channel optical nuclear potential, superscript $C(R)$ stands for the screened Coulomb potential.

To obtain the Faddeev equations in the AGS form I rewrite ${\tilde U}_{\beta\,\alpha}^{ON(R)(+)}(z)$ as
\begin{widetext}
\begin{align}
&{\tilde U}_{\beta\,\alpha}^{ONC(R)(+)}(z) = {\Delta{\overline V}_{\beta}^{ONC(R)}} +
{\Delta{\overline V}_{\beta}^{ONC(R)}}\,G^{(R)}(z)\,{\Delta{\overline V}_{\alpha}^{ONC(R)}}             \nonumber\\
&= {\Delta{\overline V}_{\beta}^{ONC(R)}} + \Big[\big({V}_{\alpha}^{N} - c_{\beta}^{\alpha}\,U_{\beta}^{ON} \big)\,G^{(R)}(z) + \big({V}_{\gamma}^{N} - c_{\beta}^{\gamma}\,U_{\beta}^{ON} \big)\,G^{(R)}(z)   + \Delta {\overline V}_{\beta}^{C(R)}\,G^{(R)}(z) \Big]\,\Delta {\overline V}_{\alpha}^{ONC(R)}       \nonumber\\
& = \Delta{\overline V}_{\beta}^{ONC(R)} + \Big[\big({V}_{\alpha}^{N} - c_{\beta}^{\alpha}\,U_{\beta}^{ON} \big){\overline G}_{\alpha}^{ONC(R)}(z)\,\big(\Delta{\overline V}_{\alpha}^{ONC(R)}\,G^{(R)}(z) + 1 \big)    \nonumber\\
&+ \big({V}_{\gamma}^{N} - c_{\beta}^{\gamma}\,U_{\beta}^{ON} \big)\,{\overline G}_{\gamma}^{ONC(R)}(z)\,\big(\Delta{\overline V}_{\gamma}^{ONC(R)}\,G^{(R)}(z)\, + 1 \big)          
 + {\Delta {\overline V}_{\beta}^{C(R)}}\,{\overline G}_{\beta}^{ONC(R)}(z)\big({\Delta{\overline V}_{\beta}^{ONC(R)}}\,G^{(R)}(z) +1 \big) \Big]{\Delta{\overline V}_{\alpha}^{ONC(R)}} 
\nonumber\\ 
&={\Delta{\overline V}_{\beta}^{ONC(R)}} + \sum\limits_{\nu}\,\Big[{\overline {\delta}_{\beta\,\nu}}\,({V}_{\nu}^{N} - c_{\beta}^{\nu}\,U_{\beta}^{(ON)}) + \delta_{\beta\,\nu}\,{\Delta {{\overline V}}^{C(R)}_{\nu}} \Big]\,{\overline G}_{\nu}^{ONC(R)}(z)\,{\tilde U}_{\nu\,\alpha}^{(R)(+)}(z),
\label{UOplus1}
\end{align}
\end{widetext}
\begin{align}
{\overline G}_{\alpha}^{ONC(R)}(z) = \frac{1}{z - T - V_{\alpha} - U_{\alpha}^{(ON)} - U_{\alpha}^{C(R)}},
\label{GalphaOR1}
\end{align}
$V_{\alpha} = V_{\alpha}^{N} + V_{\alpha}^{C(R)}$, $\,c_{\beta}^{\alpha} + c_{\beta}^{\gamma}=1$,
$\,\,\alpha \not= \beta \not = \gamma$. 

For the diagonal transition one gets from Eq. (\ref{UOplus1})
\begin{align}
&{\tilde U}_{\alpha\,\alpha}^{ONC(R)(+)}(z) ={\Delta{\overline V}_{\alpha}^{ONC(R)}} + \sum\limits_{\nu}\,\Big[{\overline {\delta}_{\alpha\,\nu}}\,({V}_{\nu}^{N} - c_{\alpha}^{\nu}\,U_{\alpha}^{(ON)})
\nonumber\\
& + \delta_{\alpha\,\nu}\,{\Delta {{\overline V}}^{C(R)}_{\nu}} \Big]\,{\overline G}_{\nu}^{ONC(R)}(z)\,{\tilde U}_{\nu\,\alpha}^{(R)(+)}(z).
\label{UOel1}
\end{align}

Then the two-particle AGS equations for the reaction amplitudes are
\begin{widetext}
\begin{align}
& {\cal T}_{\beta\,\alpha}({\rm {\bf q}}_{\beta}^{ONC(R)(-)},\,{\rm {\bf q}}_{\alpha}^{ONC(R)(+)}; E_{+})   
 = {\cal T}^{DW}_{\beta\,\alpha}({\rm {\bf q}}_{\beta}^{ONC(R)(-)},\,{\rm {\bf q}}_{\alpha}^{ONC(R)(+)}; E_{+})  \nonumber\\
&  + \sum\limits_{\nu}\int{\frac{ {\rm d} {\rm {\bf p}}_{\nu}  }{ (2\,\pi)^{3} }   } \frac{{\cal {\tilde T}}_{\beta\,\nu}^{DW}({\rm {\bf q}}_{\beta}^{ONC(R)(-)},\,{\rm {\bf p}}_{\nu}^{ONC(R)(-)}; E_{+}) \,{\cal T}_{\nu\,\alpha}({\rm {\bf p}}_{\nu}^{ONC(R)(-)},\,{\rm {\bf q}}_{\alpha}^{ONC(R)(+)}; E_{+}) }{E_{+}+\varepsilon_{\nu} - \frac{p_{\nu}^{2}}{2\,{\cal M}_{\nu}} }, 
\label{MBOa2}
\end{align}
\end{widetext}
where 
\begin{align}
&{\cal T}_{\beta\,\alpha}({\rm {\bf  q}}_{\beta}^{ONC(R)(-)},\,{\rm {\bf q}}_{\alpha}^{ONC(R)(+)};E_{+})                                                                        \nonumber\\ 
&= <\psi_{ {\rm {\bf q}}_{\beta}}^{ONC(R)(-)}\,\varphi_{\beta}^{(R)} \Big|{\tilde U}_{\beta\,\alpha}^{ONC(R)(+)} ( E_{+})\Big| \varphi_{\alpha}^{(R)}\,\psi_{ {\rm {\bf q}}_{\alpha}}^{ONC(R)(+)}>
\label{Tba1}
\end{align}
is the ONES $\alpha + (\beta\,\gamma) \to \beta + (\alpha\,\gamma)$ reaction amplitude,
\begin{align}
&{\cal T}^{DW}_{\beta\,\alpha}({\rm {\bf q}}_{\beta}^{ONC(R)(-)},\,{\rm {\bf q}}_{\alpha}^{ONC(R)(+)};E_{+})                                                      \nonumber\\                    
&= <\psi_{ {\rm {\bf q}}_{\beta}}^{ONC(R)(-)}\,\varphi_{\beta}^{(R)}\Big|{\Delta{\overline V}_{\beta}^{ONC(R)}} \Big| \varphi_{\alpha}^{(R)}\,\psi_{ {\rm {\bf q}}_{\alpha}}^{ONC(R)(+)}> 
\label{TbaDW1}
\end{align}
is the ONES post-form of the DWBA $\alpha + (\beta\,\gamma) \to \beta + (\alpha\,\gamma)$ reaction amplitude,
\begin{align}
&{\cal {\tilde T}}_{\beta\,\alpha}^{DW}({\rm {\bf q}}_{\beta}^{ONC(R)(-)},\,{\rm {\bf p}}_{\alpha}^{ONC(R)(-)};E_{+})                                                         \nonumber\\
&=<{\rm {\bf q}}_{\beta}^{ONC(R)(-)}\,\varphi_{\beta}^{(R)} \big| V_{\alpha}^{N} - c_{\beta}^{\alpha}\,U_{\beta}^{(ON)}\big| \varphi_{\alpha}^{(R)}\,{\rm {\bf p}}_{\alpha}^{ONC(R)(-)}>
\label{tildeTba1}
\end{align}
is the HOES post-form of the DWBA amplitude with the transition operator $V_{\alpha}^{N} - c_{\beta}^{\alpha}\,U_{\beta}^{(ON)}$ for the same process,
\begin{align}
&{\cal {\tilde T}}^{DW}_{\beta\,\beta}({\rm {\bf q}}_{\beta}^{ONC(R)(-)},\,{\rm {\bf p}}_{\beta}^{ONC(R)(+)};E_{+})                                                      \nonumber\\                  &= <\psi_{ {\rm {\bf q}}_{\beta}}^{ONC(R)(-)}\,\varphi_{\beta}^{(R)}\Big|{\Delta {\overline V}_{\beta}^{C(R)}} \Big| \varphi_{\beta}^{(R)}\,\psi_{ {\rm {\bf p}}_{\beta}}^{ONC(R)(+)}> 
\label{TbbDW1}
\end{align}
is the $\beta + (\alpha\,\gamma)$ elastic scattering HOES DWBA amplitude and
\begin{align}
&{\cal {\tilde T}}_{\beta\,\gamma}^{DW}({\rm {\bf q}}_{\beta}^{ONC(R)(-)},\,{\rm {\bf p}}_{\gamma}^{ONC(R)(+)};E_{+})                                                                        \nonumber\\ 
&= <\psi_{ {\rm {\bf q}}_{\beta}}^{ONC(R)(-)}\,\varphi_{\beta}^{(R)} \Big|V_{\gamma}^{N} - c_{\beta}^{\gamma}\,U_{\beta}^{(ON)} \Big| \varphi_{\gamma}^{(R)}\,\psi_{ {\rm {\bf p}}_{\gamma}}^{ONC(R)(+)}>
\label{Tbg1}
\end{align}
is the post-form of the $\gamma + (\alpha\,\beta) \to \beta +{\alpha\,\gamma}\,$ HOES DWBA reaction amplitude with the transition operator $V_{\gamma}^{N} - c_{\beta}^{\gamma}\,U_{\beta}^{(ON)}$.

Here $\psi_{ {\rm {\bf q}}_{\alpha}}^{ONC(R)(+)}$ is the distorted wave generated by the channel potential $U_{\alpha}^{ON} + U_{\alpha}^{C(R)}$. This distorted wave appears because $U_{\alpha}^{C(R)} + U_{\alpha}^{ON}$ was subtracted from ${\overline V}_{\alpha}$ in Eq. (\ref{UOplus1}).
\\

\subsection{AGS equations with optical potentials for $A(d,p)B$ reaction}
\label{AGSopticaldp1}

In this section Eq. (\ref{MBOa2}) is rewritten for the $A(d,p)B$ reaction:
\begin{widetext}
\begin{align}
& {\cal T}_{p\,A}({\rm {\bf q}}_{p}^{ONC(R)(-)},\,{\rm {\bf q}}_{A}^{ONC(R)(+)}; E_{+})   
 = {\cal T}^{DW}_{p\,A}({\rm {\bf q}}_{p}^{ONC(R)(-)},\,{\rm {\bf q}}_{A}^{ONC(R)(+)}; E_{+})  \nonumber\\
&  + \sum\limits_{i=A,p,n}\,\int{\frac{ {\rm d} {\rm {\bf p}}_{i}}{ (2\,\pi)^{3} }   } \frac{{\cal {\tilde T}}_{p\,i}^{DW}({\rm {\bf q}}_{p}^{ONC(R)(-)},\,{\rm {\bf p}}_{i}^{ONC(R)(-)}; E_{+}) \,{\cal T}_{i\,A}({\rm {\bf p}}_{i}^{ONC(R)(-)},\,{\rm {\bf q}}_{A}^{ONC(R)(+)}; E_{+}) }{E_{+} +\varepsilon_{i} - \frac{p_{i}^{2}}{2\,{\cal M}_{i}}  }.
\label{MBOa3}
\end{align}
\end{widetext}
Note that for the channel $n$  $\,\,{\rm {\bf p}}_{n}^{ON(-)}$ is the pure nuclear distorted wave because the channel Coulomb interaction $n- F$ is absent.

The effective potential  
\begin{align}
&{\cal T}^{DW}_{p\,A}({\rm {\bf q}}_{p}^{ONC(R)(-)},\,{\rm {\bf q}}_{A}^{ONC(R)(+)};E_{+})    \nonumber\\
&= <\psi_{{\rm {\bf q}}_{p}}^{ONC(R)(-)}\,\varphi_{nA} \big| V_{pn} + V_{pA} - 
U_{pB}^{ON}                                  \nonumber\\
&- U_{pB}^{C(R)} \big| \varphi_{pn}\,\psi_{{\rm {\bf q}}_{A}}^{ONC(R)(+)}>
\label{postrDWBAO1}
\end{align}
is the post form of the ONES DWBA amplitude for the $A(d,p)B$ reaction.
The other DWBA amplitudes in Eqs. (\ref{MBOa3})   are:
\begin{align}
&{\cal {\tilde T}}^{DW}_{p\,A}({\rm {\bf q}}_{p}^{ONC(R)(-)},\,{\rm {\bf p}}_{A}^{ONC(R)(-)}; E_{+})     \nonumber\\
& = <\psi_{ {\rm {\bf q}}_{p}}^{ONC(R)(-)}\,\varphi_{nA} \Big|V_{pn} \Big| \varphi_{pn}\,\psi_{ {\rm {\bf p}}_{A}}^{ONC(R)(-)}>
\label{TpAO1}
\end{align}
is the post form of the HOES DWBA amplitude for the $A(d,p)B$ reaction with the transition operator $V_{pn}$,
\begin{align}
&{\cal {\tilde T}}^{DW}_{p\,p}({\rm {\bf q}}_{p}^{ONC(R)(-)},\,{\rm {\bf p}}_{p}^{ONC(R)(-)}; E_{+})     \nonumber\\                  
& = <\psi_{ {\rm {\bf q}}_{p}}^{ONC(R)(-)}\,\varphi_{nA} \Big|V_{pA}^{C(R)} - U_{pB}^{C(R)} \Big| \varphi_{nA}\,\psi_{ {\rm {\bf p}}_{p}}^{ONC(R)(-)}>
\label{TppO1}
\end{align}
is the HOES DWBA $p + (nA)$ elastic scattering amplitude,
\begin{align}
&{\cal {\tilde T}}^{DW}_{p\,n}({\rm {\bf q}}_{p}^{ONC(R)(-)},\,{\rm {\bf p}}_{n}^{ON(-)}; E_{+})     \nonumber\\ & = <\psi_{ {\rm {\bf q}}_{p}}^{ONC(R)(-)}\,\varphi_{nA} \Big|V_{pA}^{N} - U_{pB}^{ON} \Big| \varphi_{pA}^{(R)}\,\psi_{ {\rm {\bf p}}_{n}}^{ON(-)}>
\label{TpnO1}
\end{align}
is the HOES DWBA $n + (pA) \to p + (nA)$ reaction amplitude.

I took into account that
${\Delta{\overline V}}_{p}^{C(R)} \equiv {\Delta{\overline V}}_{nA}^{C(R)} = V_{pA}^{C(R)} - U_{pB}^{C(R)}$ and  that $c_{\alpha}^{\beta} + c_{\alpha}^{\gamma}= 1$. Because there is no optical potential in the $p-n$ channel I adopt $\,c_{p}^{A}=0$ and 
$c_{p}^{n}=1$. 

Now we can analyze Eq. (\ref{MBOa3}). For the sub-Coulomb case the Coulomb distortion in the initial and final states is dominant and the proof of the peripheral character of Eq. (\ref{MBOa3}) is the same as in section \ref{AGSsepsubCoul1}, that is, the AGS amplitude of the $A(d,p)B$
reaction is well approximated by the corresponding DWBA amplitude:
\begin{align} 
&{\cal T}_{p\,A}({\rm {\bf q}}_{p}^{ONC(R)(-)},\,{\rm {\bf q}}_{A}^{ONC(R)(+)}; E_{+})  \nonumber\\  
& \approx {\cal T}^{DW}_{p\,A}({\rm {\bf q}}_{p}^{ONC(R)(-)},\,{\rm {\bf q}}_{A}^{ONC(R)(+)}; E_{+}).
\label{ONCTpADW1}
\end{align}
One important thing to note. The sub-Coulomb reaction amplitudes in subsections \ref{AGSsepsubCoul1} and \ref{SubCoulomb1} are well approximated by the sub-Coulomb DWBA amplitudes when the energies are so low that the nuclear optical  potentials can be neglected and the distorted waves in the initial and final states can be approximated by the Coulomb ones. However, when the energy, still being sub-Coulomb, increases, the approximation of the AGS reaction amplitude by the sub-Coulomb DWBA one fails. Meantime, approximation (\ref{ONCTpADW1}) works practically at all sub-Coulomb energies because the DWBA amplitude determined by Eq. (\ref{postrDWBAO1}) contains the distorted waves generated by the sum of the channel Coulomb and nuclear optical potentials. It also contains the optical potential in the transition operator.

Now I consider the $A(d,p)B$ reaction at the energies above the Coulomb barrier on heavier nuclei.
Owing to the presence of the distorted waves in the initial and final channels the DWBA amplitude can be peripheral and dominantly contributed by the tail of the of the $(nA)$ bound-state wave function. It can be easily checked using the FRESCO code \cite{FRESCO}
 Assume that it is the case and let us analyze the AGS Eq. (\ref{MBOa3}).

The first term on the  right-hand-side of this equation is the post-form of the ONES DWBA amplitude. 
Assume that it is the case and the amplitude ${\cal T}_{p\,A}^{DW}({\rm {\bf q}}_{p}^{ONC(R)(-)},\,{\rm {\bf q}}_{A}^{ONC(R)(+)}; E_{+})$ is peripheral. Its peripheral character means that it is contributed by the tail of the $(nA)$ bound-state wave function and, hence, is parametrized in terms of the ANC $C_{nA}$ of this bound state. 

The second term contains ${\cal {\tilde T}}^{DW}_{p\,A}({\rm {\bf q}}_{p}^{ONC(R)(-)},\,{\rm {\bf p}}_{A}^{ONC(R)(+)}; E_{+})$,  which is the DWBA $A(d,p)B$ reaction amplitude with the $V_{pn}$ as the transition operator. 
It is also peripheral and can be easily estimated because one can use the zero-range approximation for the $V_{pn}$. Then the radial integration in this amplitude is carried over $r_{nA}$. Since it is assumed that ${\cal T}_{p\,A}^{DW}({\rm {\bf  q}}_{p}^{ONC(R)(-)},\,{\rm {\bf q}}_{A}^{ONC(R)(+)}; E_{+})$ is peripheral, it is also true for ${\cal {\tilde T}}^{DW}_{p\,A}({\rm {\bf q}}_{p}^{ONC(R)(-)},\,{\rm {\bf p}}_{A}^{ONC(R)(+)}; E_{+})$, which is aslo is parameterized in terms of the ANC $C_{nA}$ of the $(n\,A)$ bound state.  

The third terms contains the HOES amplitude $T_{p\,A}({\rm {\bf p}}_{p}^{ONC(R)(-)},\,{\rm {\bf q}}_{A}^{ONC(R)(+)})$, which is the same reaction amplitude as the one on the left-hand-side but the HOES. 
All three first terms on the right-hand-side of Eq. (\ref{MBOa3}) provide forward peaked proton's angular distribution.
The fourth term, as in all the previous considerations, has a flat angular distribution and can be neglected compared to the first three terms when considering the angular distributions near the stripping peak.

To further simplify the AGS equation the $d+A$ elastic scattering amplitude ${\cal T}_{A\,A}({\rm {\bf p}}_{A}^{ONC(R)(-)},\,{\rm {\bf q}}_{A}^{ONC(R)(+)}; E_{+})$ is replaced by the DWBA elastic scattering amplitude 
in which the Born term is subtracted:
\begin{align}
&{\cal T}^{DW}_{A\,A}({\rm {\bf p}}_{A}^{ONC(R)(-)},\,{\rm {\bf q}}_{A}^{ONC(R)(+)};E_{+})                                                      \nonumber\\                    
&= <\psi_{ {\rm {\bf p}}_{A}}^{ONC(R)(-)}\,\varphi_{pn}\Big|{\Delta{\overline V}_{A}^{ONC(R)}} \Big| \varphi_{pn}\,\psi_{ {\rm {\bf q}}_{A}}^{ONC(R)(+)}>. 
\label{TAAoptDW1}
\end{align}
Here $\Delta{\overline V}_{A}^{ONC(R)} = \Delta{\overline V}_{pn}^{ONC(R)}= V_{pA} + V_{nA} - U_{dA}^{ON} - U_{dA}^{C(R)}$. 

Then AGS Eq. (\ref{MBOa3}) reduces to the equation 
\begin{widetext}
\begin{align}
& {\cal T}_{p\,A}({\rm {\bf q}}_{p}^{ONC(R)(-)},\,{\rm {\bf q}}_{A}^{ONC(R)(+)}; E_{+})   
 = {\cal T}^{DW}_{p\,A}({\rm {\bf q}}_{p}^{ONC(R)(-)},\,{\rm {\bf q}}_{A}^{ONC(R)(+)}; E_{+})  \nonumber\\
&  + \int{\frac{ {\rm d} {\rm {\bf p}}_{A}}{ (2\,\pi)^{3}}} \frac{{\cal {\tilde T}}_{p\,A}^{DW}({\rm {\bf q}}_{p}^{ONC(R)(-)},\,{\rm {\bf p}}_{A}^{ONC(R)(-)}; E_{+}) \,{\cal T}^{DW}_{A\,A}({\rm {\bf p}}_{A}^{ONC(R)(-)},\,{\rm {\bf q}}_{A}^{ONC(R)(+)};E_{+})  }{E_{+} +\varepsilon_{A} - \frac{p_{A}^{2}}{2\,{\cal M}_{A}}  }
\nonumber\\
&  + \int{\frac{ {\rm d} {\rm {\bf p}}_{p}}{ (2\,\pi)^{3}}} \frac{{\cal {\tilde T}}_{p\,p}^{DW}({\rm {\bf q}}_{p}^{ONC(R)(-)},\,{\rm {\bf p}}_{p}^{ONC(R)(-)}; E_{+}) \,{\cal T}_{p\,A}({\rm {\bf p}}_{p}^{ONC(R)(-)},\,{\rm {\bf q}}_{p}^{ONC(R)(+)}; E_{+}) }{E_{+} +\varepsilon_{p} - \frac{p_{p}^{2}}{2\,{\cal M}_{p}}  }
.
\label{AGSexprfin1}
\end{align}
\end{widetext}    
This is an integral equation for the $A(d,p)B$ reaction amplitude ${\cal T}_{p\,A}({\rm {\bf q}}_{p}^{ONC(R)(-)},\,{\rm {\bf q}}_{A}^{ONC(R)(+)}; E_{+})$ for the energies above the Coulomb barrier. 
I assume that that the DWBA reaction amplitude is peripheral, that is, parametrized in terms of the ANC $C_{nA}$. Hence two amplitudes on the right-hand-side of Eq. (\ref{AGSexprfin1}) are paremtrized in terms of the ANC. Then solution of this equation is also parametrized in terms of the ANC $C_{NA}$ although its dependence on the ANC may be complicated.  The more dominant contribution of the first term on the right-hand-side of Eq. (\ref{AGSexprfin1}) the closer to the linear the dependence on the ANC of its solution.   
\\

\section{Generalized AGS equations with optical potentials, three-body continuum and bound states }
\label{generalizedAGSCDCC1}

In this final section I present the generalized Faddeev equations in the two-particle AGS form, which extends the AGS equations derived in section \ref{AGsopticalpotentials1} by including three-body continuum in the spectral decomposition of the Green functions ${\overline G}_{\nu}^{ONC(R)}$
and a few bound states. The three-body continuum can be taken into account using the CDCC  method \cite{Austern,thompson}. The three-body wave function in this approach
takes the form \cite{thompson}
\begin{align}
\Psi^{CDCC(+)}_{\alpha} = \sum\limits_{i=0}^{N}\,{\overline {\varphi}}_{\alpha(i)}\,\psi_{\alpha(i)}^{(+)}, 
\label{PsiCDCC1}
\end{align}
where $\Psi^{CDCC}_{\alpha}$ is the CDCC wave function in the channel $\alpha$, $\,{\overline {\varphi}}_{\alpha(i)}$ is the internal wave function of the couple $\alpha$, $\,\psi_{\alpha(i)}$
is the wave function of the relative motion of particle $\alpha$ and the pair $(\beta\,\gamma)$. The sum over $i$ includes the sum over bound states and discretized continuum states of the pair $\alpha$.  
The wave function ${\overline {\varphi}}_{\alpha(i)}$ is the normalized bin function for the discretized continuum state $\,i\,$ or the normalized bound-state wave function corresponding to the relative energy of the pair $\,\alpha$ $\,\,{\overline E}_{\alpha(i)}\,$ ($\,{\overline E}_{\alpha (i)}\,$ is positive for continuum bins and negative for bound states). To calculate the radial part of $\,{\overline {\varphi}}_{\alpha(i)}\,$ for the continuum states the bins are used (for the details see \cite{thompson}).
The wave function $\,\psi_{\alpha(i)}\,$ is the solution of the Schr\"odinger equation in the potential $\,U_{\alpha}^{C(R)} + U_{\alpha}^{ON}\,$ with energy $\,E - {\overline E}_{\alpha(i)},\,$ where $\,E\,$ is the total energy of the three-body system. 
   
The spectral decomposition of the Green function is given by
\begin{widetext}
\begin{align}
{\overline G}_{\alpha}^{(R)}(z) = \sum\limits_{i}^{N}\,\int \frac{ {\rm d} {\rm {\bf p}}_{\alpha} }{(2\,\pi)^{3}}\, \frac{|{\overline {\varphi}}_{\alpha(i)}^{(R)}\,\psi_{{\rm {\bf p}}_{\alpha}}^{ONC(R)(-)}><{\overline {\varphi}}_{\alpha\,(i)}^{(R)}\,\psi_{{\rm {\bf p}}_{\alpha}}^{ONC(R)(-)}|}{z - 
{\overline E}_{\alpha(i)} -  \frac{p_{\alpha}^{2}}{2\,{\cal M}_{\alpha}} },
\label{GnuCDCC1}
\end{align}
\end{widetext}
where $\,{\overline E}_{\alpha\,(i)}>0\,$ is the relative energy of the pair $\,\alpha\,$ for the discretized continuum state $i$ (center energy of the bin $i$) and $\,-\varepsilon_{\alpha\,(i)}\,$ for the bound state $i$.
Note that in this spectral decomposition the energy $\,\frac{p_{\alpha}^{2}}{2\,{\cal M}_{\alpha}}\,$ of the relative motion of particle $\alpha$ and the bound state $(\beta\,\gamma)$ and ${\overline E}_{\alpha(i)}$ are independent.

Then generalizing of the AGS equations (\ref{MBOa2}) by including the discretized continuum and bound states we get
\begin{widetext}
\begin{align}
& {\cal T}_{\beta (i)\,\alpha (j)}({\rm {\bf q}}_{\beta}^{ONC(R)(-)},\,{\rm {\bf q}}_{\alpha }^{ONC(R)(+)})   
 = {\cal T}^{DW}_{\beta(i)\,\alpha(j)}({\rm {\bf q}}_{\beta}^{ONC(R)(-)},\,{\rm {\bf q}}_{\alpha}^{ONC(R)(+)})  \nonumber\\
&  + \sum\limits_{s}\,\int{\frac{ {\rm d} {\rm {\bf p}}_{\alpha}  }{ (2\,\pi)^{3} }   } \frac{{\cal {\tilde T}}_{\beta (i)\,\alpha (s)}^{DW}({\rm {\bf q}}_{\beta }^{ONC(R)(-)},\,{\rm {\bf p}}_{\alpha }^{ONC(R)(-)}) \,{\cal T}_{\alpha (s)\,\alpha (j)}({\rm {\bf p}}_{\alpha }^{ONC(R)(-)},\,{\rm {\bf q }}_{\alpha }^{ONC(R)(+)}) }{E- {\overline E}_{\alpha(s)} - \frac{p_{\alpha}^{2}}{2\,{\cal M}_{\alpha}} + i0 }            \nonumber\\
& + \sum\limits_{s}\int{\frac{ {\rm d} {\rm {\bf p}}_{\beta}  }{ (2\,\pi)^{3} }   } \frac{{\cal {\tilde T}}_{\beta(i)\,\beta(s)}^{DW}({\rm {\bf q}}_{\beta}^{ONC(R)(-)},\,{\rm {\bf p}}_{\beta}^{ONC(R)(-)}) \, {\cal T}_{\beta(s)\,\alpha(j)}({\rm {\bf p}}_{\beta}^{ONC(R)(-)},\,{\rm {\bf q}}_{\alpha}^{ONC(R)(+)}) }{E- {\overline E}_{\beta(s)} - \frac{p_{\beta}^{2}}{2\,{\cal M}_{\beta}} + i0 }              \nonumber\\
& + \sum\limits_{s}\int{\frac{ {\rm d} {\rm {\bf p}}_{\gamma}  }{ (2\,\pi)^{3} }   } \frac{{\cal {\tilde T}}_{\beta(i)\,\gamma(s)}^{DW}({\rm {\bf q}}_{\beta}^{ONC(R)(-)},\,{\rm {\bf p}}_{\gamma}^{ONC(R)(-)}) \,{\cal T}_{\gamma(s)\,\alpha(j)}({\rm {\bf p}}_{\gamma}^{ONC(R)(-)},\,{\rm {\bf q}}_{\alpha}^{ONC(R)(+)}) }{E -{\overline E}_{\gamma(s)} - \frac{p_{\gamma}^{2}}{2\,{\cal M}_{\gamma}} + i0 }. 
\label{MBOCDCC1}
\end{align}
\end{widetext}
All the matrix elements are defined by equations given in section \ref{AGsopticalpotentials1} 
in which the bound state wave functions $\varphi_{\nu}$ should be replaced 
by ${\overline {\varphi}}_{\nu(i)}$.

\section{Summary}

Usually, for the analysis of the $(d,p)$ reactions the DWBA, ADWA or CDCC methods \cite{Austern,Johnson,thompson}are being used.
In these last two approaches the coupling of the neutron transfer channel with the deuteron breakup channel is taken effectively into account, while the explicit coupling to the proton and heavy-particle transfer channels 
and elastic scattering is neglected. Meantime, the Faddeev equations allow us to take into account the coupling of all the transfer, elastic and breakup channels simultaneously. 
In this paper I am formulating the formalism of the three-body Faddeev equations for the $(d,p)$ reactions using the two-body AGS equations. For separable potentials these equations are exact and can be used for the analysis 
of the direct $A(d,p)B$ reactions on heavier nuclei at sub-Coulomb energies.
The advantage of the AGS equations with separable potentials is that the effective potentials are given by a few simple diagrams. The sum of the pole and triangle exchange diagrams can be expressed in terms of the DWBA amplitude for the sub-Coulomb $(d,p)$ reactions. 
 For local potentials to obtain the two-body AGS equations I neglect the contribution from the deuteron breakup channel taking into account explicitly the coupling to transfer and elastic scattering channels. For low-energy reactions, especially for the sub-Coluomb ones, the contribution from the breakup channel is small and the developed formalism is well suited for the  direct sub-Coulomb $A(d,p)B$ reactions on heavier nuclei. It is shown that the AGS equation for the sub-Coulomb $A(d,p)B$ reactions are peripheral and dominated by the post-form of the DWBA amplitude, which is peripheral. Hence, the AGS amplitude is also parametrized in terms of the ANC.

In this paper the two-body AGS equations are also generalized by including the optical potentials in the same manner as it is done in the DWBA. Naturally, the effective potentials in the obtained AGS equations are the DWBA amplitudes. 
Although it is shown that the AGS $A(d,p)B$ reaction amplitude can be parametrized in terms of the ANC $C_{nA}$ of the bound state $(nA)$, there is a conceptual problem of determination of the ANC from comparison of the AGS cross section with experimental data. The problem is that the AGS equations are based on the three-body model. Hence the AGS amplitude contains only the single-particle $(nA)$ bound-state wave function rather than the overlap integral, which includes the spectroscopic factor. I will address this issue in the following up paper.
Finally, in this work the two-body AGS equations with optical potentials are generalized by including the intermediate three-body continuum and more than one bound state in each channel.

\section{Acknowledgments}
This work was supported by the U.S. DOE Grant No. DE-FG02-93ER40773, NNSA Grant No. DE-NA0003841
and U.S. NSF Award No. PHY-1415656.

\end{document}